\documentclass[12pt,preprint]{aastex}

\usepackage{enumerate}
\usepackage{url}
\usepackage{epstopdf}
\shorttitle{Variable Evolved Stars}
\shortauthors{Polsdofer et al.}

\begin{document}


\title{Examining the Infrared Variable Star Population Discovered in the Small Magellanic Cloud Using the SAGE-SMC Survey}

\author{
Elizabeth Polsdofer\altaffilmark{1,2*}
, J. Seale\altaffilmark{3}
, M. Sewi{\l}o \altaffilmark{4,3}
, U. P. Vijh\altaffilmark{5}
, M. Meixner\altaffilmark{2,3}
, M. Marengo\altaffilmark{1}
, M. Terrazas\altaffilmark{5*,6}
}

\altaffiltext{1}{Iowa State University, Department of Physics and Astronomy, 12 Physics Hall, Ames, Iowa 50011, USA}
\altaffiltext{2}{Space Telescope Science Institute, 3700 San Martin Dr., Baltimore, MD 21218, USA}
\altaffiltext{3}{The Johns Hopkins University, Department of Physics and Astronomy, 366 Bloomberg Center, 3400 N. Charles Street, Baltimore, MD 21218, USA} 
\altaffiltext{4}{Space Science Institute, 4750 Walnut Street, Suite 205, Boulder, CO 80301, USA}
\altaffiltext{5}{Ritter Astrophysical Research Center, University of
Toledo, Toledo, OH 43606, USA}
\altaffiltext{6}{Washington State University, Department of Physics and Astronomy, Pullman, WA 99164, USA}
\altaffiltext{*}{Summer intern}
\email{empolsdofer@gmail.com}
\begin{abstract}

We present our study on the infrared variability of point sources in the Small Magellanic Cloud (SMC). We use the data from the {\it Spitzer} Space Telescope Legacy Program ``Surveying the Agents of Galaxy Evolution in the Tidally Stripped, Low Metallicity Small Magellanic Cloud'' (SAGE-SMC) and the ``{\it Spitzer} Survey of the Small Magellanic Cloud'' (S$^{3}$MC) survey, over three different epochs, separated by several months to three years. Variability in the thermal infrared is identified using a combination of {\it Spitzer}`s IRAC 3.6, 4.5, 5.8, and 8.0 $\mu$m bands, and the MIPS 24 $\mu$m band. An error-weighted flux difference between each pair of three epochs (``variability index'') is used to assess the variability of each source. A visual source inspection is used to validate the photometry and image quality. Out of $\sim$2 million sources in the SAGE-SMC catalog, 814 meet our variability criteria. We matched the list of variable star candidates to the catalogs of SMC sources classified with other methods, available in the literature. Carbon-rich Asymptotic Giant Branch (AGB) stars make up the majority (61\%) of our variable sources, with about a third of all of our sources being classified as extreme AGB stars. We find a small, but significant population of oxygen-rich AGB (8.6\%), Red Supergiant (2.8\%), and Red Giant Branch ($<$1\%) stars. Other matches to the literature include Cepheid variable stars (8.6\%), early-type stars (2.8\%), young-stellar objects (5.8\%), and background galaxies (1.2\%). We found a candidate OH maser star, SSTISAGE1C J005212.88-730852.8, which is a variable O-rich AGB star, and would be the first OH/IR star in the SMC, if confirmed. We measured the infrared variability of a rare RV Tau variable (a post-AGB star) that has recently left the AGB phase.  Fifty nine variable stars from our list remain unclassified. 

\end{abstract}


\section{Introduction}
\label{s:intro}

Variability is a common feature of many astronomical objects, among which are pulsating stars, disks around Young-Stellar Objects (YSOs), accreting binary systems, and Active Galactic Nuclei (AGNs). The processes behind variability phenomenon are diverse: they range from sporadic changes in the accretion rate of YSO and AGN disks (see \citealp{flaherty12} and \citealp{peterson00}, respectively) to the periodic variations of the opacity and ionization states in pulsating stars' interiors (\citealt{ita11, matsunaga11, madore09}). Monitoring the brightness variations induced by these processes is a powerful probe to understanding the underlying physical nature of sources.

Systematic long term variability monitoring has been conducted at visible wavelengths since the beginning of the 20$^{\rm th}$ Century, especially following the discovery of the Cepheid Period-Luminosity relation \citep{leavitt12}. More recently, the advent of space-based infrared observatories extended these studies to the thermal infrared. At thermal infrared wavelengths emission is mainly due to warm astronomical dust heated by stellar sources or accretion processes. Infrared variability studies offer a unique window to characterize the temporal variations in stellar mass loss and accretion. Indeed, data from the InfraRed Astronomical Satellite (IRAS) and the Infrared Space Observatory (ISO), have been used to correlate the pulsation properties and mass loss rates of evolved stars (e.g.,  \citealt{vanderveen90, omont99, whitelock91}), as well as the properties of disk accretion in optically variable YSOs (e.g.,  \citealt{juhasz07, abraham04}). These pioneering studies were, however, limited to bright isolated objects in the Galaxy due to the low sensitivity and spatial resolution of the available thermal infrared instrumentation.

The \emph{Spitzer} Space Telescope \citep{werner04} and the Wide-field Infrared Survey Explorer (WISE; \citealt{wright10}) for the first time provided the sensitivity required to repeatedly survey large areas of the sky to register the temporal variations of mid-IR sources. The observing strategy of the WISE mission, in particular, allowed the satellite to scan across each object in the sky at least 12 times every 90 minutes. The time series photometry obtained by this observatory are available as part of the recently released AllWISE catalog and have been used to characterize the light curves of short period variables sources in the Galaxy (e.g.,  \citealt{klein14}). \emph{Spitzer}, which operates as a facility telescope, allows monitoring with arbitrary timescales, constrained only by on-sky visibility as the observatory orbits around the Sun. \emph{Spitzer's} superior sensitivity and angular resolution (as good as $\sim$2 arcsec, a factor of three better than WISE) opened the possibility of monitoring the infrared variability of stars and disks in crowded areas like star forming regions in the plane of the Galaxy (e.g.,  \citealt{giannini09}), globular clusters in the Galactic halo, and the stellar population of the Galactic bulge (e.g.,  the ongoing {\it Spitzer} Carnegie RR Lyrae Program, monitoring stellar variability in 31 Galactic clusters and the bulge). Implementing \emph{Spitzer} as a research telescope also rendered Local Group galaxies accessible for variability studies at infrared wavelengths.

The {\it Spitzer} Legacy Programs ``Surveying the Agents of a Galaxy's Evolution'' (SAGE; \citealt{meixner06}) and ``Surveying the Agents of Galaxy Evolution in the Tidally Stripped, Low Metallicity Small Magellanic Cloud'' (SAGE-SMC; \citealt{gordon11a}), in particular, are ideally suited for studying the infrared variable population in the Magellanic Clouds, two of the closest Milky Way satellites. The scientific goals of these two surveys focus on the life cycle of baryonic matter as traced by dust emission, from its start in the Interstellar Medium (ISM) where new stars form, to the chemical enrichment of the ISM by the dusty winds of evolved stars. The two programs imaged the Magellanic Clouds with both the InfraRed Array Camera (IRAC; \citealt{fazio04}) 3.6, 4.5, 5.8, and 8.0 $\mu$m bands, and the Multiband Imaging Photometer for \emph{Spitzer} (MIPS; \citealt{rieke04}) 24, 70, and 160 $\mu$m bands.  Each galaxy was mapped twice in those surveys, allowing the detection of infrared variability over timescales roughly commensurable to the separation (several months to three years) between the two epochs. An analysis of the SAGE Large Magellanic Cloud (LMC) data was presented in \citet{vijh09}. Using the IRAC 3.6--8.0 $\mu$m and MIPS 24 $\mu$m data they found $\sim $2000 variable sources (out of over 4 million point sources in the SAGE catalog), the majority of which ($\sim$87\%) were identified as evolved stars in the Asymptotic Giant Branch (AGB) phase. A small population of variable YSO candidates was also detected by \citet{vijh09}.

In this work, we focus on the SMC, following a similar approach to that developed by \citet{vijh09}. The SMC, with a distance of $\sim$61 kpc \citep{haschke12, hilditch05}, is characterized by an even lower metallicity than the LMC ($\sim$1/5 solar vs. $\sim$1/2 solar; \citealt{russell92, rolleston99, rolleston03, lee05}), below the threshold, where the properties of the ISM are expected to drastically change, as traced by the low dust mass fraction \citep{engelbracht05, draine07, sandstrom10}. The morphology of the SMC is characterized by a pronounced ``Bar'' with a ``Wing'', and an extended ``Tail'' pointing toward the LMC (see Figure \ref{f:fig1}). The latter structure is part of what is generally known as the Magellanic Bridge, the H\,{\sc i} filament connecting SMC with the LMC (e.g., \citealt{mcgee86}). The Magellanic Bridge is a clear manifestation of the tidal interaction between the two galaxies and of a possible close encounter $\sim$200~Myr ago \citep{zaritsky04, harris07}. The SAGE-SMC observations covered the full SMC including the Bar, Wing, and the Tail. We have supplemented this data with the pathfinder ``{\it Spitzer} Survey of the Small Magellanic Cloud'' (S$^3$MC; \citealt{bolatto07}), to provide a third epoch photometry for the portion of the SMC Bar and Wing. While the SMC has been extensively monitored for optical variability as part of the MAssive Compact Halo Objects Project (MACHO; \citealp{alcock96}) and Optical Gravitational Lensing Experiment (OGLE-III; \citealp{udalski08}) microlensing experiments, this work is the first extensive survey of the SMC searching for variability at thermal infrared wavelengths.

This paper is organized as follows. In Section~\ref{s:thedata}, we describe the {\it Spitzer} SAGE-SMC and S$^3$MC point source catalogs, as well as other optical and infrared data we used to identify and characterize the SMC variables. Section~\ref{s:variablesstaridentification} outlines our methods for selecting and validating candidate variable sources. In Section~\ref{s:classification}, we compare the list of SMC variables to SMC sources classified with other methods, available in the literature. The demographics of the variables identified in this study is then discussed in Section~\ref{s:discussion}, and a summary of our findings is given in Section~\ref{s:conclusions}.

\section{The Data}
\label{s:thedata}

\begin{figure}[h!]
  \centering
      \includegraphics[scale=.75]{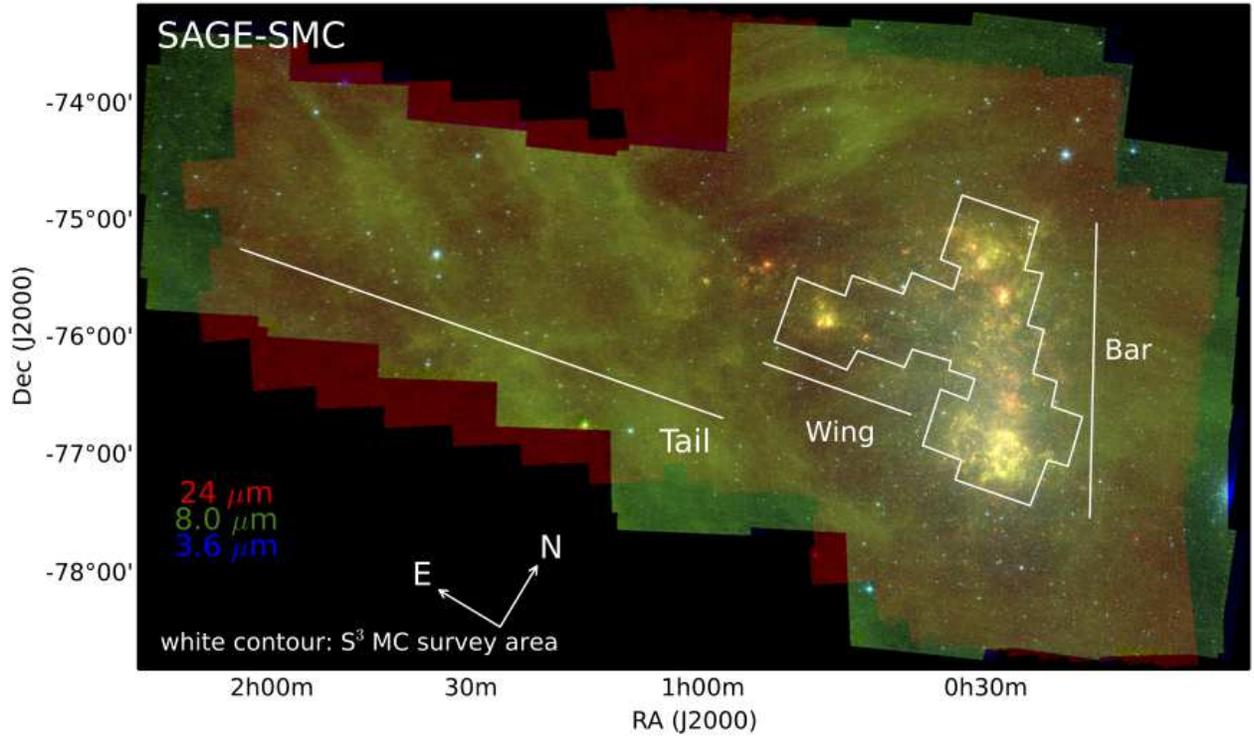}
  \caption{The three-color composite image of the SMC combining the SAGE-SMC Epoch 1 MIPS 24 $\mu$m (red), IRAC 4.5 $\mu$m (green), and IRAC 3.6 $\mu$m (blue) images. The white contour outlines the S$^{3}$MC survey area. The main components of the SMC (Bar, Wing, and Tail) are indicated.}
\label{f:fig1}
\end{figure}

We utilized the data taken in three separate epochs with \textit{Spitzer} during the cryogenic mission using the IRAC 3.6, 4.5, 5.8, and 8.0 $\mu$m bands, and the MIPS 24~$\mu$m band. As mentioned before, the data were collected as part of the SAGE-SMC and S$^3$MC surveys:

\begin{enumerate}[1.]

\item The Surveying the Agents of Galaxy Evolution in the Tidally Stripped, Low Metallicity Small Magellanic Cloud (SAGE-SMC; \citealt{gordon11a}) mapped the SMC in all IRAC (3.6, 4.5, 5.8, and 8.0 $\mu$m) and MIPS (24, 70, and 160 $\mu$m) bands in two separate epochs, which we identify as Epoch 1 and Epoch 2 for the remainder of this paper. These two epochs are separated by $\sim$3~months (IRAC, data obtained in June and September 2008) and $\sim$9~months (MIPS, acquired in September 2007 and June 2008), respectively. The observations covered the full SMC, including the Bar, Wing, and part of the Magellanic Bridge (the Tail; see Figure~\ref{f:fig1}), for a total coverage of $\sim $30~deg$^2$. 

\item The {\it Spitzer} Survey of the Small Magellanic Cloud (S$^{3}$MC; \citealt{bolatto07}) only mapped the Bar and Wing of the SMC ($\sim$8 times smaller coverage than SAGE-SMC, for a total of $\sim $4~deg$^2$; see white contour in Figure~\ref{f:fig1}) in one single epoch (November 2004 for MIPS and May 2005 for IRAC), preceding the SAGE-SMC survey by about 3 years. The S$^{3}$MC IRAC (3.6--8.0 $\mu$m) and MIPS (70--160 $\mu$m) data were reprocessed using the SAGE-SMC pipeline as Epoch 0 data, providing a consistently reduced data set for all three epochs. In the following we refer to the reprocessed S$^3$MC data as Epoch 0.

\end{enumerate}


The SAGE-SMC survey provides two catalogs containing the sources' photometry for individual epochs, i.e. the IRAC \emph{Catalog} and the IRAC \emph{Archive} (see \citealt{gordondoc}). For our study, we selected the more reliable Catalog over the Archive, favoring detection reliability and photometric accuracy over completeness. Similarly, we adopted the more restrictive MIPS 24~$\mu$m Catalog over the more complete, but less reliable Full List \citep{gordondoc}.  The IRAC Catalog contains $\sim$217,000, $\sim$1.23 million, and $\sim$1.13 million sources in Epoch 0, Epoch 1, and Epoch 2, respectively. The MIPS 24~$\mu$m Catalog includes $\sim$5,600 Epoch~0, and $\sim$16,500 Epoch 1 and 2 sources. We did not use photometry from the MIPS 70 and 160 $\mu$m bands due to the lower spatial resolution ($\sim$18$''$ and $\sim$40$''$, respectively) relative to the shorter wavelength data. The IRAC and MIPS 24 $\mu$m data have angular resolution of 1$\rlap.{''}$7--2$''$ and 6$''$, respectively. The Epoch 0 IRAC photometry has a limiting sensitivity of 0.045, 0.028, 0.12, and 0.10~mJy in the IRAC 3.6, 4.5, 5.8, and 8.0~\micron{} bands, respectively. The sensitivity of the Epoch 1 and 2 is about a factor of two lower due to the shorter integration time. The MIPS 24~\micron{} sensitivity is instead similar in the three epochs, and is $\sim$0.7~mJy.

In addition to the source catalogs, our study also uses the SAGE-SMC Epoch 0--2 IRAC and MIPS 24 $\mu$m images for source quality validation.

The MIPS data were consistently taken before the IRAC data with a period of no less than three months between observations. The MIPS and IRAC data were taken at different times within each epoch, which can allow us to use the data from MIPS to confirm a source detected as variable based on the IRAC data on a different temporal cadence. Table \ref{t:tab1} provides information about when the data were taken by IRAC and MIPS for each epoch. The longest time between observations with IRAC and MIPS is nine months in Epoch 1, contrasted to the shortest time of three months in Epoch 2. Epoch 0 IRAC and MIPS data were taken six months apart from each other, but the time gap between Epochs 0 and 1 is nearly three years.

The time interval between the different measurements is the main limiting factor to the typical period of the sources that we are able to detect in our dataset. The detectability of a source as a variable will depend on how well the sampling of the light curve matches the period of variability: our study is most sensitive to sources which our instruments are able to detect in different phases of their variability period when we collect our data. These types of sources include long period variables that have not completed a full phase of their variability period during our study. Sources that are in the same phase of their variability period when our data were collected, will not be detected as variable. As a consequence, our study is biased toward redder sources with a long period of variability (of the orders of months to years) with high amplitudes of variations in the thermal infrared.

\subsection{Ancillary Data: Optical and Near-IR Point Source Catalogs}
\label{s:ancillarydata}

The near-IR $JHK_s$ data from the combined Two Micron All-Sky (2MASS; \citealt{skrutskie06}) and the 2MASS 6X Deep (6X2MASS; \citealt{cutri04}) Point Source Catalogs are the integral part of the SAGE-SMC IRAC catalogs (see \citealt{gordondoc}).  The SAGE-SMC catalog is cross-matched in the SAGE-SMC database with the optical data from the Magellanic Clouds Photometric Survey (MCPS, {\it UBVI} bands; \citealp{zaritsky02}) and the Optical Gravitational Lensing Experiment (OGLE-III, {\it VI} bands; \citealp{udalski08}).  We considered IRAC and optical sources a match when a distance between their positions was $\leq$1$''$.  

The ancillary data were used to construct sources' spectral energy distributions for a wide wavelength range, from optical to 24 $\mu$m. We adopt the  {\it V}- and {\it I}-band photometry from OGLE-III when available, from MCPS otherwise.

\section{Variable Star Identification}
\label{s:variablesstaridentification}

We identified the SMC variable sources following the procedure described in \citet{vijh09} for the LMC. We first matched the individual SAGE-SMC  Epoch 0, Epoch 1, and Epoch 2 catalogs to extract a time series for each source detected in all epochs. We then applied variability criteria based on the flux change between epochs to select sources with measurable change above the photometric sensitivity. We finally performed quality checks aimed to remove spurious detections in order to have a reliable catalog of SMC infrared variables.

\subsection{Interepoch Matching}
\label{s:interepochmatching}

In the initial step of our variable stars selection process, we cross-matched single epoch IRAC and MIPS 24 $\mu$m catalogs. Accurate matching between epochs is crucial to ensure that we are measuring changes in flux for the same sources across multiple epochs. The match was performed using the CasJobs interface within the Mikulski Archive for Space Telescopes (MAST). CasJobs is based on a SQL database.

We started by generating a list of all the sources that had coordinates within 1$''$ of each other within at least two of the three epochs of the SAGE-SMC IRAC data. We then matched the resulting IRAC lists with their corresponding MIPS 24 $\mu$m data for each epoch, again requiring a match within 1$''$. The requirement to identify common sources across multiple epochs of data significantly reduced the number of sources available for our study. For instance, the largest of our matched lists, containing sources with matches in Epoch 1 and 2 (the largest spatial coverage of the observations), contains just over 800,000 sources with valid 3.6~\micron{} photometry, compared to over 2 million sources in each single epoch (Epoch 1 and Epoch 2) catalog. The source counts for other IRAC and MIPS 24 $\mu$m bands and  pairs of epochs can be found in Table~\ref{t:tab2}. Note that our method of searching for variables would miss novae and supernovae if the progenitor star was not detectable in our survey before the outburst.

\subsection{Variability Criteria}
\label{s:variabilitycriteria}

We identified candidate variables out of our inter--epoch matched catalogs on the basis of the variability index $V$ defined by \cite{vijh09} as the error-weighted flux difference between each pair of epochs:

\begin{equation}
V = \frac{f_i - f_j}{\sqrt{\sigma_{f_i}^2 + \sigma_{f_j}^2}}
\end{equation}

\noindent
where $f_i$ and $f_j$ are fluxes in a certain band for a single source in epochs $i$ and $j$, respectively; and $\sigma_{f_i}$ and $\sigma_{f_j}$ are respective flux uncertainties. This index was calculated for each possible pair of the three available epochs ($i,j = 0, 1, 2$ and $i \ne j$) in each band, so that a single source can have up to 15 variability indices if it has valid fluxes in all the IRAC and MIPS 24 $\mu$m bands across all three epochs of data. Note that a positive $V$  would correspond to an increase in flux between the two epochs, while a negative index would indicate that the source decreased in flux from one epoch to the other.

We considered a source to be variable if $|V| > 3$ (equivalent to measuring flux changes with at least $3 \sigma$ sensitivity) in at least two neighboring bands, again following \citet{vijh09}:

\begin{equation}
\left\{
\begin{array}{lll}
&&|V_{3.6_{ij}}| > 3 \textrm{~~~and~~~} |V_{4.5_{ij}}| > 3\\
&&|V_{4.5_{ij}}| > 3 \textrm{~~~and~~~} |V_{5.8_{ij}}| > 3\\
&&|V_{5.8_{ij}}| > 3 \textrm{~~~and~~~} |V_{8.0_{ij}}| > 3\\
&&|V_{8.0_{ij}}| > 3 \textrm{~~~and~~~} |V_{24_{ij}}| > 3
\end{array}
\right. \textrm{~~~with~~~} i \ne j \textrm{~~~and~~~} i,j = 0, 1, 2
\end{equation}

\noindent
where $i$ and $j$ are two of the three available epochs. We also required that the variability index among neighboring IRAC bands have the same sign, since all the IRAC bands were observed at essentially the same time within each epoch and we expect the change from one epoch to another to be similar for each band. Failing this condition is an indication that the flux in different bands may come from mismatched sources, sources whose flux is contaminated by a nearby or blended source, or by the presence of an artifact in our data. The exception to this rule is for the 24 $\mu$m photometry, for which we do not require the same sign variability; the MIPS data were taken at different times than the IRAC data and correspond to a different variability phase.  We finally excluded all sources with magnitudes larger than 15 at 3.6~\micron{} and 4.5~\micron{}, 13 at 5.8~\micron{}, and 12 in the 8.0~\micron{} IRAC bands; the photometric uncertainties increase for fainter sources, making flux measurements less reliable \citep{gordondoc}.

Note that to be considered as variable star candidates, sources do not need to be variable between all three pairs of epochs (only $\sim$6\% of all candidate variable sources have $|V| > 3$ in all three epoch pairs). Sources do not need to have valid photometry in necessarily all IRAC and MIPS bands as long as the sources have enough valid fluxes to demonstrate their variability. 

\begin{figure}[h!]
\centering
      \includegraphics[scale=.6]{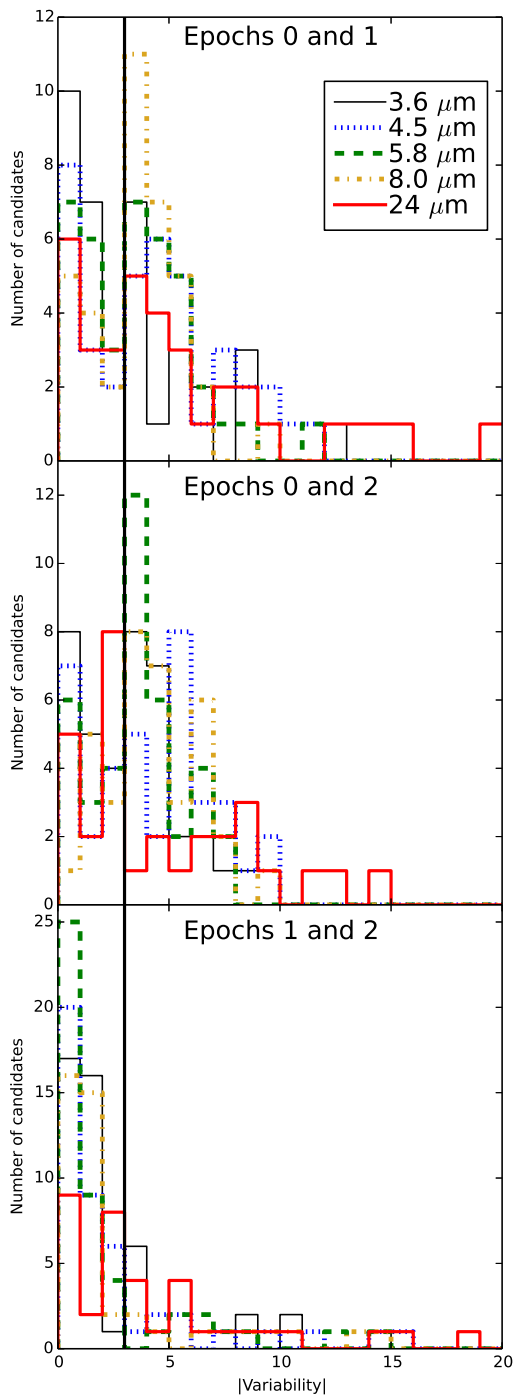}
  \caption{Distribution of the absolute values of variability indices ($|V|$) in each IRAC and MIPS 24 $\mu$m band, calculated for the interval between Epochs 0 and 1 (top), 0 and 2 (middle), and 1 and 2 (bottom). Measurements between Epochs 0 and 2 yielded a similar spread of $|V|$ as between Epochs 0 and 1; both sample variability over a period of a few years. Measurements between Epochs 1 and 2 instead sample a change over a period of just a few months. Sources with  $|V|>$ 3 in at least two consecutive bands are classified as variable sources.}
\label{f:fig2}
\end{figure}

The results of our search for variable stars is summarized in Table \ref{t:tab2} and the distributions of the variability indices for all bands and epoch pairs are shown in Figure~\ref{f:fig2}. The variability indices for IRAC bands between Epochs 0 and 1 and Epochs 0 and 2 have a larger width (as shown in Figure~\ref{f:fig2}) than the variability indices between Epochs 1 and 2. This is a consequence of the longer integration time used in the S$^3$MC survey (reprocessed into SAGE-SMC Epoch 0), leading to smaller flux uncertainty and hence a better sensitivity for  flux changes measured by the variability index. Conversely, the uncertainly in the MIPS 24~\micron{} photometry (and hence the spread in the variability index distribution) was similar between the SAGE-SMC and S$^3$MC datasets. We found 825 sources that meet our criteria for variability in at least two consecutive bands and pairs of epochs.

\subsection{Source Inspection}
\label{s:sourceinspection}

We have visually inspected each variable star candidate in all available bands and epochs using thumbnail images extracted from the SAGE-SMC IRAC 3.6, 4.5, 5.8, and 8.0 $\mu$m, and MIPS 24 $\mu$m band mosaics. The purpose of the visual inspection was to remove candidates whose photometric reliability was compromised by poor pixel quality, optical or electronic artifacts, un-rejected transients such as cosmic rays, or contamination from a neighboring source. At the end of this process we were left with 814 candidates that meet all our criteria for variability, photometry, and image quality.

\subsection{The SMC Infrared Variables Catalog}
\label{s:catalog}

Our final catalog of infrared variables in the SMC is available in full within the online version of this paper. The catalog lists the basic astrometric and photometric properties of each variable candidate, including IRAC and MIPS designations, fluxes and magnitudes with uncertainties in the optical (MCPS and OGLE-III), near-IR (2MASS/6X2MASS), and mid-IR (\emph{Spitzer} IRAC and MIPS 24~\micron) bands, as well as the variability indices $V$ for each pair of epochs. The catalog also includes possible classification of each source based on matches with the literature (see Section~\ref{s:classification} below). The complete list of columns in the catalog of SMC variables is provided in Table~\ref{t:tab3}.

\section{Source Classification}
\label{s:classification}

In order to understand the demographics of 814 newly-identified variable star candidates, we have cross-matched our source list with a number of published catalogs of SMC sources.  These catalogs were constructed using different methods and include sources classified with varying levels of confidence.  We required a distance between a variable source and its matching source from literature to be $\leq$1$''$.  A detailed discussion of the catalogs we used to determine the nature of variable sources and the matching results follows. The summary is provided in Table~\ref{t:tab4}.

For several SAGE-SMC variable sources we found multiple matches in literature. Resolving ambiguity in source classification is out of scope of this paper, thus we provide all available classifications in our catalog (columns 19 to 21 in Table~\ref{t:tab3}).

\subsection{Evolved Stars}
\label{ss:evolvedstars}

Over 80\% of the LMC infrared variables identified by \citet{vijh09} are evolved stars: a similar result is to be expected in our SMC search. Stars of all masses, once they leave the main sequence, enter one or more phases when their convective envelopes expand by several orders of magnitude, while their effective temperatures become as low as $\sim$3,000~K. In those conditions, these stars cross the Long Period Variables (LPVs) instability strip in the Hertzsprung-Russell diagram, where radial pulsations are modulated by periodic changes in the opacity and ionization of internal stellar layers. These pulsations are observed as luminosity variations with period ranging from several tens to thousands of days. According to the shapes, amplitudes, and periods of their optical light curves, LPVs are classified as Mira, Semiregular, or Irregular variables \citep{samus12}, with Miras typically having larger amplitude luminosity variations. LPVs are also characterized by intense mass loss, which leads to the formation of dusty circumstellar envelopes. The circumstellar envelopes in turn give rise to strong infrared excesses above the photospheric continuum. While the variability properties of LPVs are well characterized in the visible, their mid-infrared variability has only recently been studied in a systematic way (e.g.,  \citealt{riebel10} and references therein).

Low and intermediate mass stars ($M \la 8 \ M_\odot$) will become LPVs twice in their life: as stars in the Red Giant Branch (RGB, characterized by H nuclear burning in a shell around an inert He core), and in the Asymptotic Giant Branch (AGB, characterized by alternate H and He burning in a shell around an inert C/O core). While the AGB phase is shorter than the RGB phase, AGB stars tend to be more luminous, show larger variability, and have more intense mass loss. As a consequence, our search is biased toward discovering AGB rather than RGB variables. AGB stars are subjected to \emph{third dredge-up} events, in which elements synthesized during He-shell burning events are transported toward the stellar surface. The end product is the injection of trace elements (the so-called $s$-elements) in the stellar atmosphere, uniquely produced during these He-shell burning events. Third dredge-up events ultimately lead to a gradual increase in the photospheric C/O abundance ratio that, for a star with a main sequence mass in the $\sim1.5$ -- $4 \ M_\odot$ range (for solar metallicity; \citealt{straniero95, straniero97}), leads to the formation of a so-called \emph{carbon star}. Carbon stars are characterized by a C-rich atmospheric and circumstellar dust and gas chemistry. Stars below this mass limit never experience enough third dredge-up episodes to become C-rich AGBs, while stars above the upper limit destroy the newly generated C in a process called Hot Bottom Burning (HBB; \citealt{smith85, boothroyd92}). Stars that undergo HBB processes remain O-rich AGBs and tend to have much higher luminosities than low mass O-rich AGBs. Note that in the low metallicity environments, like in the Magellanic Clouds, stars have a lower initial O atmosphere abundance, and a smaller number of dredge-up events are required to push the C/O ratio above unity. In this case the lower mass limit for the product of C-rich AGBs is somewhat reduced and a larger population of C-rich AGBs is to be expected \citep{ferrarotti06, ventura12}.

The mass loss rate dramatically increases toward the end of the AGB phase \citep{willson00}; in this late stage, the circumstellar envelope of AGB stars becomes optically thick and completely obscures the star in optical wavelengths. Stars of this kind are often referred to as extreme AGBs and can be either O-rich (in which case they may show OH/IR maser activity) or C-rich. A handful of the most extreme AGB stars are responsible for the larger fraction of dust injection into the LMC and SMC ISM \citep{boyer12, riebel12}, which these dust contributors tending to be prevalently C-rich \citep{boyer11}.

High mass stars, on the other hand, cross the LPV instability strip as Red SuperGiant (RSG) stars. RSG stars are characterized by a higher luminosity than low mass AGB stars (but comparable to intermediate mass HBB O-rich AGB stars) and a similar mass loss rate. RSG stars tend to have much smaller populations than AGB and RGB stars due to the Initial Mass Function being biased toward lower masses, and the shorter life span of RSG stars. As such, RSG stars are not large contributors to the ISM and we should not expect to find many of them among our candidate variables.

We matched our SMC variable sources with the color-selected evolved stars from \citet{boyer11} and with the LPVs identified in the OGLE-III project by \citet{soszynski11} based on the characteristics of their light curves (and then classified as O-rich or C-rich on the basis of their optical and near-IR colors). \citet{boyer11} differentiate the stars according to the evolutionary phases (RGBs, O-rich AGBs, C-rich AGBs, extreme AGBs, and RSGs), while \citet{soszynski11} only differentiate between RGBs, O-rich AGBs, and C-rich AGBs (along with providing details on the variability type). We also matched our sample of SAGE-SMC variables with the catalog of massive stars from \citet{bonanos10} that includes spectroscopically identified RSG stars. As expected, most of the matches with our variable candidates are AGB stars, of which 82 are O-rich and 497 are C-rich AGBs. Of those O- and C-rich AGBs, 251 are classified as extreme AGBs according to \citet{boyer11}, the vast majority of which (85\%) are C-rich according to \citet{soszynski11}. For the non-extreme AGBs, \citet{boyer11} and \citet{soszynski11} are generally in agreement, with $\sim$16\% mismatch within the O-rich category and less than 3\% disagreement for the carbon stars. We only found 6 RGB and 23 RSG stars in our study, but it is worth noting that the \citet{boyer11} classification scheme cannot distinguish between RGB stars and AGB stars with luminosity below the tip of the RGB.

\citet{boyer11} has also three categories that we have not considered in our analysis. Sources labeled as ``Far InfraRed objects'' (FIR) are characterized by increasing flux at 8 and 24~\micron, even larger than what is expected for extreme AGB stars. The fact that the FIR sources are mostly found in proximity of star forming region supports the hypothesis that they may be misclassified YSOs rather than evolved stars, but without spectroscopic analysis it is not possible to rule out the possibility that our FIR sources are evolved stars with unusually thick dusty envelopes. For this reason we have left FIR sources as unclassified in our survey. The ``anomalous'' aO-rich and aC-rich AGBs are O-rich and C-rich star, respectively, with peculiar colors in the \citet{boyer11} analysis. Rather than keeping these intermediate classes, we have relied on the chemical classification provided by the OGLE-III analysis in \citet{soszynski11}, which is  based on a combination of optical and near-IR colors. We found that most (18 out of 27) aO-rich AGBs are confidently described as O-rich AGBs in the OGLE-III catalog, with the remaining being classified as C-rich AGB stars, while the two aC-rich AGBs are evenly split between C- and O-rich AGBs in OGLE-III.

Due to their lower brightness, we were unable to calculate the $V_{24}$ index of RGB stars for any of our intervals of variability. RGB stars also have $|V_{8.0}| < 3$ across all variability intervals. One source, SSTISAGE1C J004906.89-724309.7, has contrasting classifications among different catalogs. \cite{soszynski11} lists the source as a RGB star with O-rich chemistry and \cite{boyer11} classifies the source as a C-rich AGB star. This source is the only RGB candidate we found that varies in the 5.8 $\mu$m band (all other RGB stars show variability in the 3.6  and 4.5~\micron{} IRAC bands and have $|V_{58}| < 3$ within all epochs).

Two RSGs have contrasting matches in literature. SSTISAGE1C J010302.41-720152.9 is classified as RSG by \cite{bonanos10} and as O-rich AGB star by \citet{boyer11}. Given the higher reliability of the \cite{bonanos10} classification that used full spectra rather than broadband colors,  SSTISAGE1C J010302.41-720152.9 is likely to be an under-luminous RSG star. This source meets our variability criteria primarily in the IRAC bands between Epochs 0 and 1, but the MIPS 24~\micron{} band has the largest variability of all RSG sources with $V_{24} = -15.80$ (between Epochs 0 and 2) and $-12.92$ (between Epochs 1 and 2). Note, however, that \citet{soszynski11} is not attempting to differentiate AGB from RSG stars. RSG stars appear in the OGLE-III catalog as Semiregular LPVs with O-rich chemistry. This suggests that this source may be a RSG based on the $K_s$ luminosity cut in \citet{boyer11}. Interestingly, this source is the most variable source between Epochs 1 and 2 in terms of consecutive neighboring bands with $|V| > 3$.

Finally, we have identified a candidate planetary nebula (PN), matched from a catalog kindly provided by G. Jacoby (2014, private communication). However, this same source (SSTISAGE1C J005530.37-725021.9) is also cross-matched with a YSO from \citet{bolatto07} and a background AGN from \citet{kozlowski11}. The variability of this source is detected only between Epochs 0 and 1 and Epochs 0 and 2 in the 5.8 and 8.0 $\mu$m bands. This suggests that this object may be changing brightness on very long time scales (comparable to the $\sim$3~years separation between the S$^3$MC and SAGE-SMC surveys), pointing toward changes in accretion rate as a possible mechanism for the infrared variability.

\subsection{Cepheids}
\label{ss:cepheids}

Intermediate mass stars (4--$8 \ M_\odot$) cross the Cepheid instability strip during the He-core burning stage, while they describe a pronounced ``blue loop'' on the Hertzsprung-Russell diagram. Cepheids are radial pulsators similar to LPVs, but their typical variability period is between 1 and 80 days. Cepheids have F and G spectral types; typically they do not have substantial mass loss (but see \citealt{marengo10, barmby11, matthews12}) and their intrinsic brightness makes them easily detectable in the SAGE-SMC survey. Given their importance as distance indicators, thanks to their period-luminosity relation (Leavitt law, \citealt{leavitt12}), Cepheid variables have been extensively studied not just in the Milky Way, but also in local group galaxies (including the SMC) at visible wavelengths (e.g., \citealt{freedman01}), as well as in the infrared \citep{ngeow08, ngeow10, scowcroft11, scowcroft13}.

To search for Cepheids, we matched our catalog of candidate variables with the source lists in the OGLE-III database (where, again, the classification is based on the characteristics of the light curves). We matched both Classical Cepheids \citep{soszynski10a} and the lower mass Type II Cepheids \citep{soszynski10b}. We found 70 Cepheids in our catalog (9\% of our sample), with 67 of sources being classified as Classical Cepheids, and the remaining 3 being classified as Type II Cepheids.  Our Cepheids show the highest variability index in the 3.6 and 4.5~\micron{} bands, but generally have $|V|< 5$ across all intervals of variability in all bands.

We also matched our catalog with the list of OGLE-III RR Lyrae in the SMC from \citet{soszynski10c}. We did not find any matches, most likely due to the lower amplitudes and luminosities that RR Lyrae variables have in the thermal infrared \citep{madore13}.

\subsection{Early Type Stars}
\label{ss:earlytype}

Early type stars (high mass stars with typically O and B spectral types) are characterized by several different kinds of variability, in some cases due to periodic pulsations (e.g., radially pulsating $\beta$~Cephei and non-radially pulsating $\alpha$ Cygni stars), in others due to episodic outbursts (e.g., Luminous Blue Variables, whose prototype is $\beta$~Dor in the LMC). In other cases, variability in early type stars is caused by accretion in binary systems and can manifest itself at wavelengths ranging from the X-rays to the infrared (e.g., X-ray binaries). Early type stars tend to be rare, even more so the ones that show detectable variability.

To search for early type variables in the SMC, we have cross matched our variable star candidates list with the catalog of spectroscopically confirmed massive stars from \citet{bonanos10} and dusty OB stars from \citet{sheets13}. We found a total of 23 sources ($\sim$3\% of our sample), 16 of which are early B stars, and one is also classified as a ``dusty OB star'' in \citet{sheets13}. Four of the early spectral type stars identified by \citet{bonanos10} are also matched with the \citet{sturm13} catalog of SMC X-ray sources. Those sources are likely accreting X-ray binaries.

\subsection{Young Stellar Objects}
\label{ss:yso}

According to the \citet{hartmann85} model of accretion outbursts in FU Orionis objects, infrared variability in YSOs is generally associated with changes in the accretion rate modulated by the protostellar disk rather than precessing jets \citep{herbig03}, which are rather invoked to explain optical variability. 

Observations at different wavelength regimes allow a detection of YSOs at different stages of their evolution. The SED of a YSO changes when it evolves. Stage 0 protostars are in the early protostellar collapse stage with the majority of the source mass residing in the infalling envelope. The SED is dominated by the far-IR, because the protostar and possibly a disk are deeply embedded in a circumstellar envelope that absorbs the forming star's photospheric emission, warms the dust and re-emits it in the far-IR.  Stage I is a later stage of protostellar collapse. Stage I YSOs display very broad SEDs that peak near 100 $\mu$m; they emit the bulk of their radiation at the mid-IR to far-IR wavelengths. The envelope mass is similar to the mass of the central protostellar object. Stage I YSOs have well-developed accretion disks and their envelopes have bipolar cavities excavated by outflows.  Stage II YSOs are characterized by the presence of optically thick disks, bipolar outflows, and possibly the remains of a tenuous infalling envelope. Stage II YSOs emit mostly in the near-IR to mid-IR.  Stage III YSOs, which are mostly detected at optical to near-IR wavelengths, have optically thin disks or no disks, and may still have optical bipolar jets. 

The mid-IR \emph{Spitzer} data are sensitive to Stage I and Stage II YSOs. Prior to \emph{Spitzer}, there was only one YSO identified in the SMC \citep{gatley82}. In the galaxy-wide study based on the \emph{Spitzer} SAGE-SMC data, \citet{sewilo13} identified $\sim$1000 YSO candidates which are categorized into ``possible YSO candidates'' and ``high-reliability YSO candidates.''  An earlier study by \citet{bolatto07} that used the S$^{3}$MC data (and thus concentrated on the main body of the SMC) found  $\sim$300 YSO candidates; this list is less vetted than that of \citet{sewilo13} and potentially more contaminated by non-YSOs. \citet{chen14} used the SAGE-SMC data to conduct a photometric search for YSOs in the SMC Tail and identified 26 massive YSO candidates.  More detailed studies on YSOs in the SMC concentrated on individual star-forming regions: NGC\,346 \citep{simon07} and NGC\,602 \citep{carlson07,carlson11, gouliermis07}, identifying $\sim$150 additional YSO candidates; their YSO lists include low luminosity sources missed by galaxy-wide studies.  \citet{sewilo13} used the relatively conservative brightness cuts to remove typically dimmer background galaxies from their YSO list, and thus selecting the brightest, most massive YSOs.  The detailed inspection of source morphology that would help to distinguish between background galaxies and YSOs can only be done on smaller scales. To date, \emph{Spitzer} studies identified over 1,200 individual YSO candidates in the SMC; 33 sources are confirmed spectroscopically as bona fide YSOs 
\citep{vanloon08,vanloon10,oliveira11,oliveira13}.   

Operating at longer wavelengths, the \emph{Herschel Space Observatory} \citep{pilbratt10} is sensitive to lower dust temperatures than \emph{Spitzer}, allowing a detection of YSOs at the earliest evolutionary stages that had been missed by \emph{Spitzer}.  \emph{Herschel} is sensitive to Stage 0 and Stage I YSOs. \citet{seale14} used the data from the \emph{Herschel} Key Project HERITAGE (``Herschel Inventory of the Agents of Galaxy Evolution''; \citealt{meixner13}), which imaged the Magellanic System at 100, 160, 250, 350, and 500 $\mu$m, to identify populations of sources in the LMC and SMC. In the SMC, \citet{seale14} identified $\sim$700 YSO candidates, which are categorized into ``probable YSO candidates'' and ``possible YSO candidates''; 24\% of these are newly-identified YSO candidates.  

We cross–matched our list of candidate SMC variables with the YSO candidate lists from literature to identify infrared variable YSOs.  We found 
44 matches in \citet{sewilo13},
13 in \citet{bolatto07}, 
10 in \citet{seale14},
4 in \citet{oliveira13}, 
3 in \citet{carlson11}, and 
one match in each \citet{vanloon08}, \citet{vanloon10}, and \citet{oliveira11} (all 3 refer to the same source).  Many sources appear on more than one YSO list from literature; our list of candidate SMC variables includes 47 individual YSO candidates.

However, out of 47 sources matched to YSO candidates, 11 were also matched to other classes of sources. Five sources were matched to extreme AGB stars \citep{boyer11}, tending to have very high variability indices at the 24 $\mu$m band ($|V_{24}| \ga 20$) between Epochs 1 and 2.  The remaining sources were matched to the early type stars from \citet{bonanos10}, and the multiple-match PN described in Section~\ref{ss:evolvedstars} above. 

Of the 984 YSOs in \citet{sewilo13}, we match 44, suggesting a lower limit of $\sim$4.5\% SMC YSOs that are variable in the infrared with our sensitivity and timescale coverage.  \citet{sewilo13} estimated physical parameters of 22 of these sources based on the results of the SED fitting with the YSO radiative transfer models of \citep{robitaille06}. All 22 sources are classified as Stage I YSOs, indicating that the infrared variability may be tracing the accretion variability.

\subsection{Background Galaxies}
\label{ss:galaxies}

The OGLE-III survey has been used to search for optical variability of AGNs and quasars (QSOs) as part of the Magellanic Quasars Survey \citep{kozlowski11,kozlowski13}. These sources are clearly background objects and not part of the SMC stellar population. Still, finding background galaxies within our survey is a useful complement to AGN and QSO optical studies, which typically show flux changes between $\sim$3--30\% in the visible, over a period of 10--100~days \citep{kelly09}. The detection of variability at \emph{Spitzer} wavelengths would indicate the presence of a non-thermal component in the characteristic active galaxies' mid-IR spectral ``bump'' (e.g.,  \citealt{impey88, neugerbauer99}, and references therein). Furthermore, identification of distant quasars projected within the field of view of the SMC can be a useful tool to determine the proper motion of this galaxy.

We matched our catalog with  \citet{kozlowski11, kozlowski13} lists of AGNs and QSOs, which have been identified through spectroscopy and light curve morphology collected over a time span of 12 years. We found 8 sources matching the \citet{kozlowski11} AGN list, and one matching the \citet{kozlowski13} QSO sample. The latter source is also matched to an X-ray source from the \citet{sturm13} catalog (SSTISAGE1C J011139.57-725031.6). As mentioned above, one of the variable sources matched to the AGN (SSTISAGE1C J005530.37-725021.9) is also matched to a PN (G. Jacoby 2014, private communication), and a YSO (Bolatto et al. 2007).

\subsection{Other Catalogs}
\label{ss:other}

Finally, we have also matched our sample of variable candidates with other source lists with peculiar spectral properties, namely $\sim$3,000 X-ray sources from \citet{sturm13} and a catalog of radio sources from \citet{wong12} (457 sources detected at 3~cm and 601 sources at 6~cm). We found a match for 6 X-ray sources, the majority of which are associated with early type stars (hence most likely X-ray binaries), and one (SSTISAGE1C J011139.57-725031.6) in common with \citet{kozlowski13} QSOs. We only identified one 6~cm radio source as an infrared variable (less than 1\% of \citealt{wong12} catalog); this variable \emph{Spitzer} source (SSTISAGE1C J005212.88-730852.8) is also matched to the evolved stars from the \citet{boyer11} and \citet{soszynski11} catalogs (see discussion in Section~\ref{ss:ohirstar}).

\subsection{Unclassified Objects}
\label{ss:unclassified}

At the end of our classification process, we remained with 59 sources ($\sim$7\% of the total) that are not matched to any of the catalogs described above. These unclassified sources tend to be fainter (by $\sim$1 magnitude on average) than the classified variables and lack 24~\micron{} photometry (less than 20\% of the unclassified sources have 24~\micron{} photometry in at least two epochs). The unclassified sources that do have at least one epoch detection at 24~\micron{} in most cases possess an SED resembling the one of YSOs (see discussion in Section~\ref{ss:sed}). 

Many of the unclassified sources are located in the SMC Tail (see Section~\ref{s:discussion}), as expected since the Magellanic Bridge is not covered by many of the surveys we used for source classification.

\section{Discussion}
\label{s:discussion}

\begin{figure}[h!]
  \centering
      \includegraphics[scale=.55]{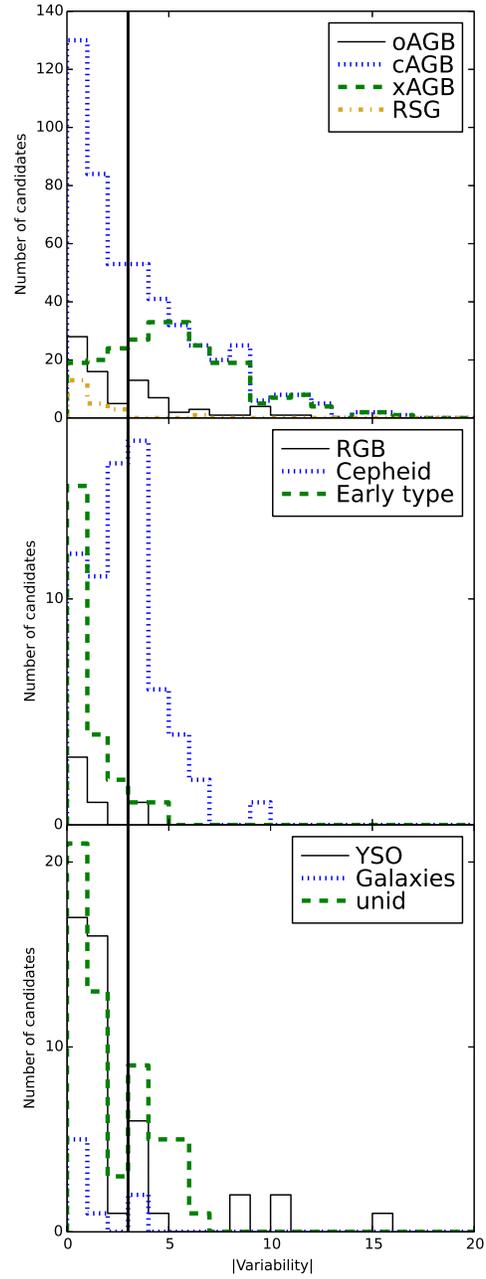}
  \caption{Distribution of the absolute values of variability indices ($|V|$) between Epochs 1 and 2 for the IRAC 3.6 $\mu$m band for several classes of sources as indicated in the legends. The largest variability indices are found for Cepheids and extreme AGB stars. The vertical line corresponds to $|V|$=3.}
\label{f:fig3}
\end{figure}

Figure~\ref{f:fig3} shows the distribution of the absolute values of variability indices in the 3.6~\micron{} band between Epochs 1 and 2, separately for each class of sources. We chose this wavelength and pair of epochs because this is the combination providing the largest number of sources for which a variability index is available (mainly due to the higher sensitivity of this band and full coverage of the SMC in Epochs 1 and 2). The histograms for other bands and combinations of epochs are similar.

Figure~\ref{f:fig3} shows that most of the sources identified as variables according to the criteria delineated in Section~\ref{s:variabilitycriteria}, have a 3.6~\micron{} variability index below the minimum threshold of 3. The fact that these sources are still classified as variables indicates that high variability is still detected in at least two other bands or in another pair of epochs. The exception are extreme AGBs with an average variability of 5.06, well above the threshold, and Cepheids, that have a median absolute variability index of 2.85, still significantly higher than other classes. It should be noted that these sources are the ones with the longest (extreme AGBs) and shortest (Cepheids) periods of variability among the type of variables we matched: this increases the chances of detection by our method by reducing the possibility of temporal aliases with the inter-epoch matching in our datasets.

\begin{figure}[htb]
\centering
  \begin{tabular}{@{}ccc@{}}
    \includegraphics[width=.95\textwidth]{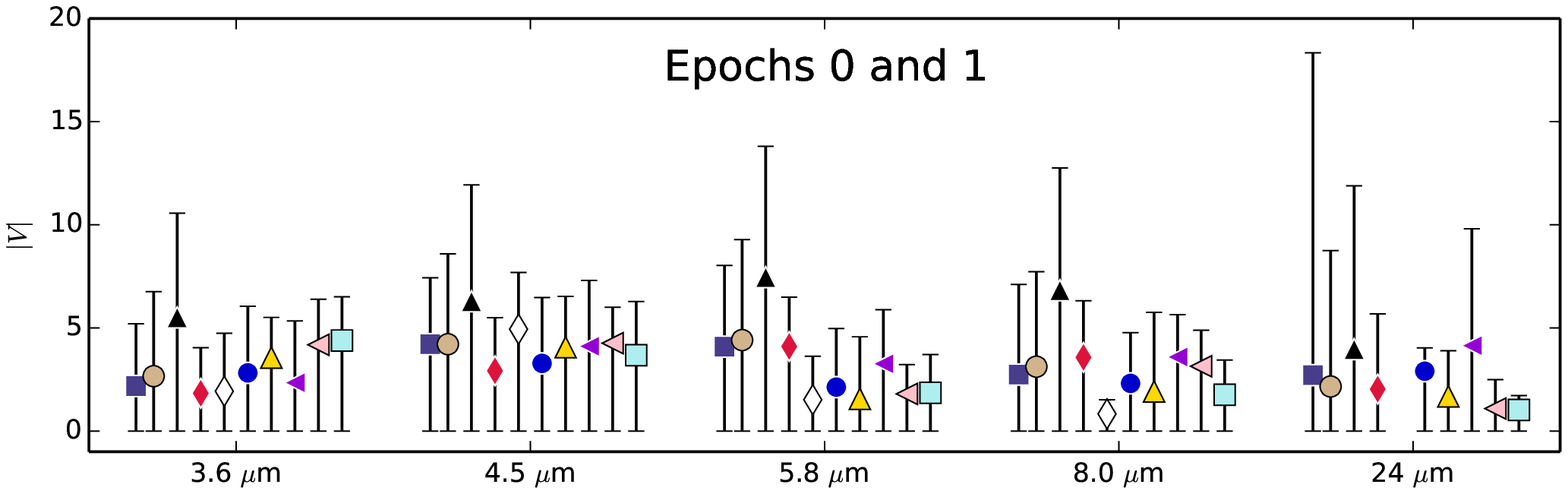} \\
    \includegraphics[width=.95\textwidth]{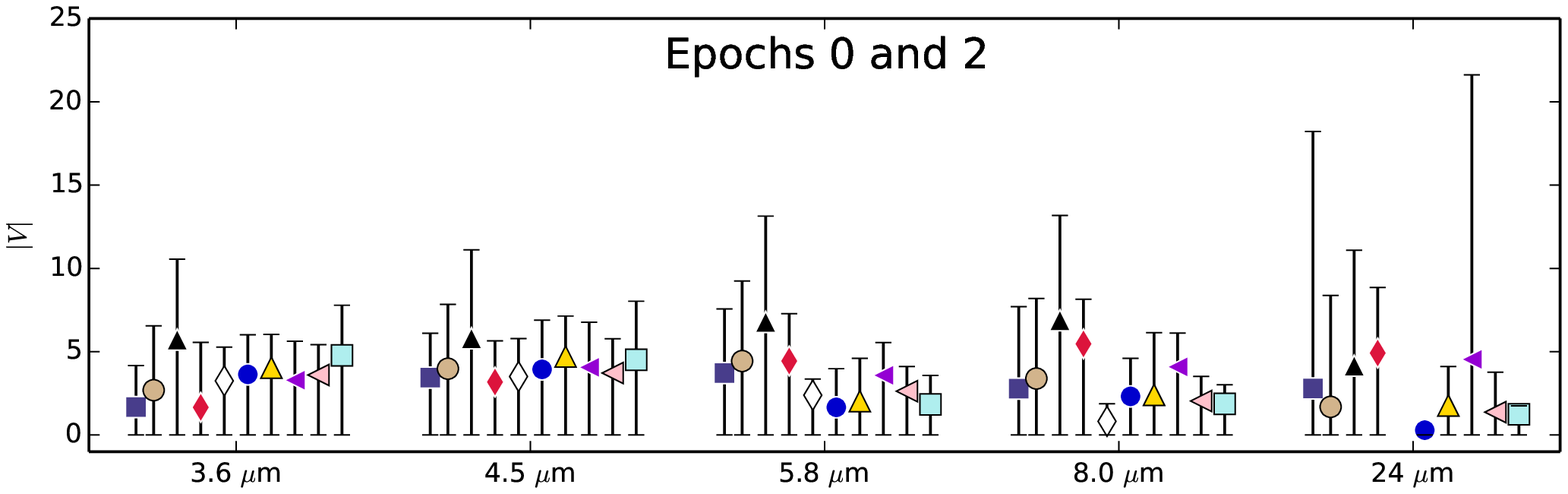} \\
    \includegraphics[width=.95\textwidth]{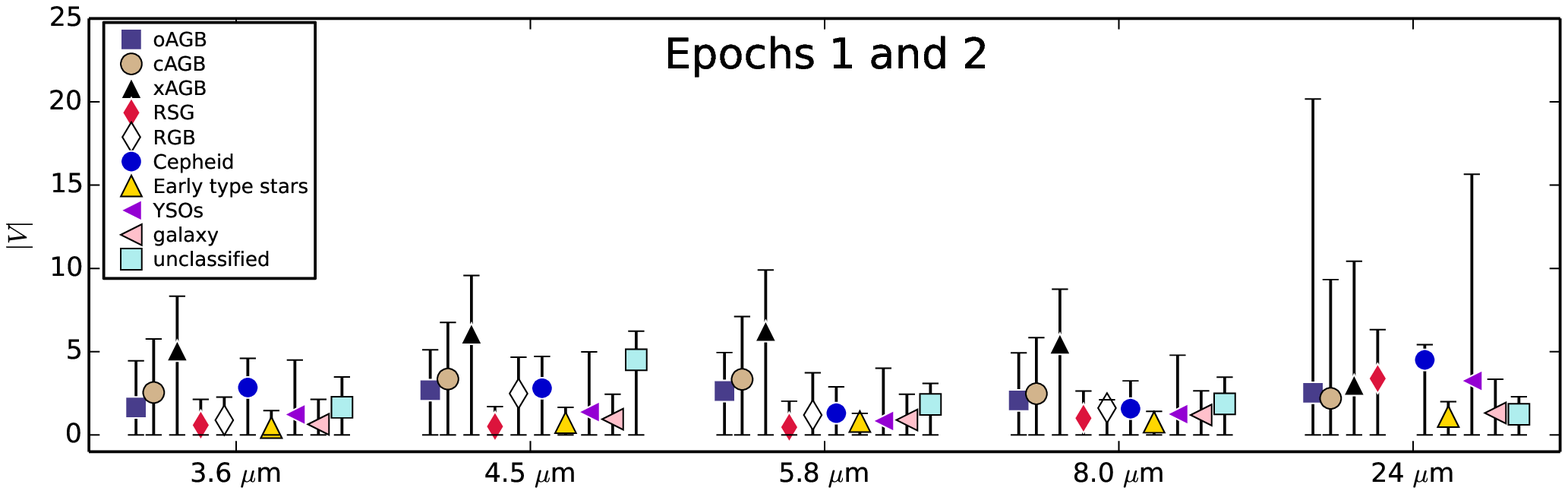}
  \end{tabular}
  \caption{Median value of the absolute variability index ($|V|$) for each band, each pair of three epochs, and each source class. The standard deviations of the variability index distributions are represented by vertical bars. The variability indices are larger for the longer inter-epoch interval, sampling time-scales of $\sim$3 years, with the exception of Cepheids that have similar variability in all intervals. O-rich AGBs and YSOs have the largest spread in $|V|$ at 24 $\mu$m, while C-rich AGBs have the largest spread in all bands.}
\label{f:fig4}
\end{figure}

A different way to look at the same behaviors is shown in Figure~\ref{f:fig4}. This figure plots the median value of the absolute variability index $|V|$ for each class, each band, and each pair of epochs, separately. The vertical bars represent the standard deviation in the distribution of $|V|$. The variability indices tend to be larger when calculated with respect to Epoch 0, in part because this epoch corresponds to the more sensitive observations obtained with the S$^3$MC survey. The much shorter time interval between the two SAGE-SMC epochs (Epochs 1 and 2), however, does play a role for some classes of sources. Cepheid stars, for example, show almost the same variability index between Epochs 1 and 2 as in the other two intervals, possibly because of their very short period which is equally sampled in all three intervals. RSGs, early type stars, YSOs, and galaxies, on the other hand, show very little variability between Epochs 1 and 2, even in bands (3.6 and 4.5~\micron{} in particular) where they have high variability in the other two intervals (Epochs 0 and 1 and Epochs 0 and 2). O-rich AGBs and YSOs have the largest spread in variability indices in the 24~\micron band; in all other bands the sources showing the largest standard deviation are instead the extreme AGB stars.

\begin{figure}[h!]
  \centering
      \includegraphics[scale=.75]{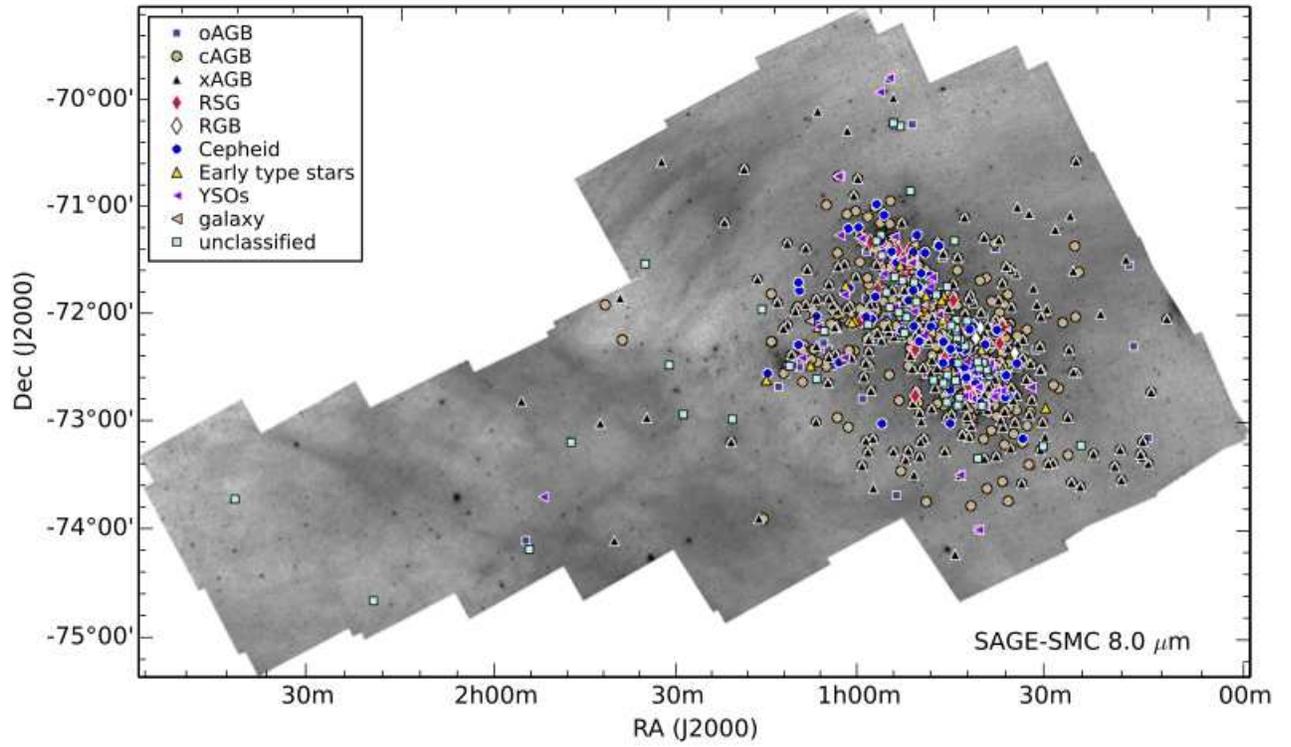}
  \caption{The spatial distribution of the variable star candidates in the SMC. The sources are overlaid on the SAGE-SMC 8.0 $\mu$m image. Sources with multiple matches have all classifications plotted.}
\label{f:fig5}
\end{figure}

Figure~\ref{f:fig5} shows the spatial distribution of the individual variables over the SAGE-SMC 8.0~\micron{} map. The figure shows that most of the variables are concentrated in the SMC Bar and Wing, due to the larger stellar density in these parts of the galaxy. Young sources (YSOs and moderately massive objects such as early type stars, RSGs, and Cepheids) are over-represented in the Bar, which is where most star formation is observed to occur. However, one YSO is detected in the Tail (SSTISAGE1C J014915.82-743225.5), which indicates the presence of active star formation sites in the bridge connecting the LMC with the SMC, as suggested by \citet{harris07} and recently confirmed by \citet{chen14}. Evolved stars are found at larger distances from the Bar, including the Tail, as expected for their lower mass and longer life span, which allows them to diffuse farther from the areas where they are born. Note a large number of symbols for extreme AGB stars overlapping with C-rich AGBs, which indicates that the extreme AGB stars in the SMC have in most cases a carbon-rich chemistry (at least in the areas of the Bar and Wing, where OGLE-III survey data are available). Finally, it should be noted that a large fraction of the variable sources in the SMC Tail are unclassified: this is a consequence of the lack of coverage in that area in the various source catalogs that we have used for source classification.

\subsection{Color-Magnitude Diagrams}
\label{ssec:cmd}

\begin{figure}
        \centering
           \begin{tabular}{@{}ccc@{}}
                \includegraphics[width=.5\textwidth]{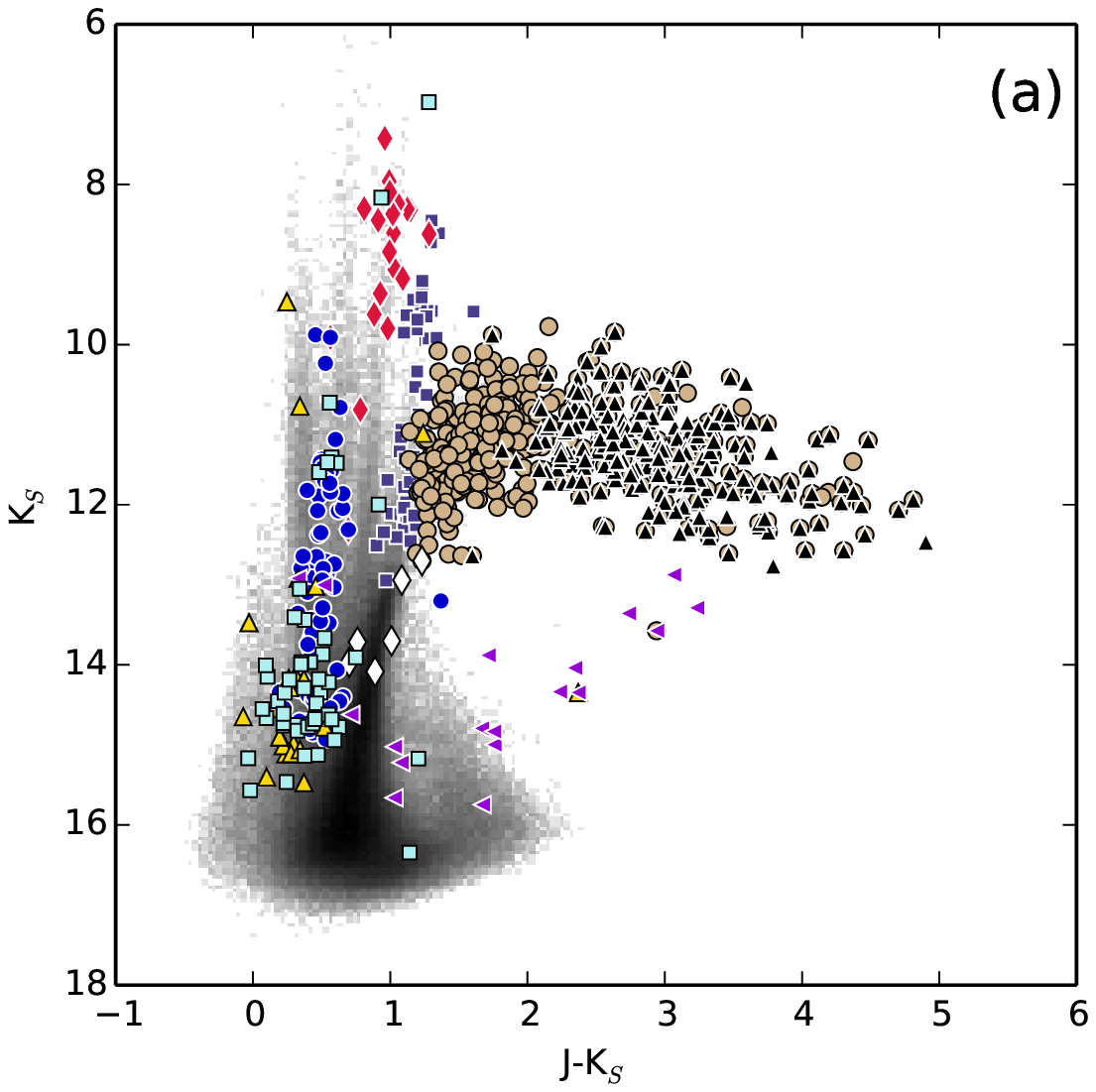}

                \includegraphics[width=.5\textwidth]{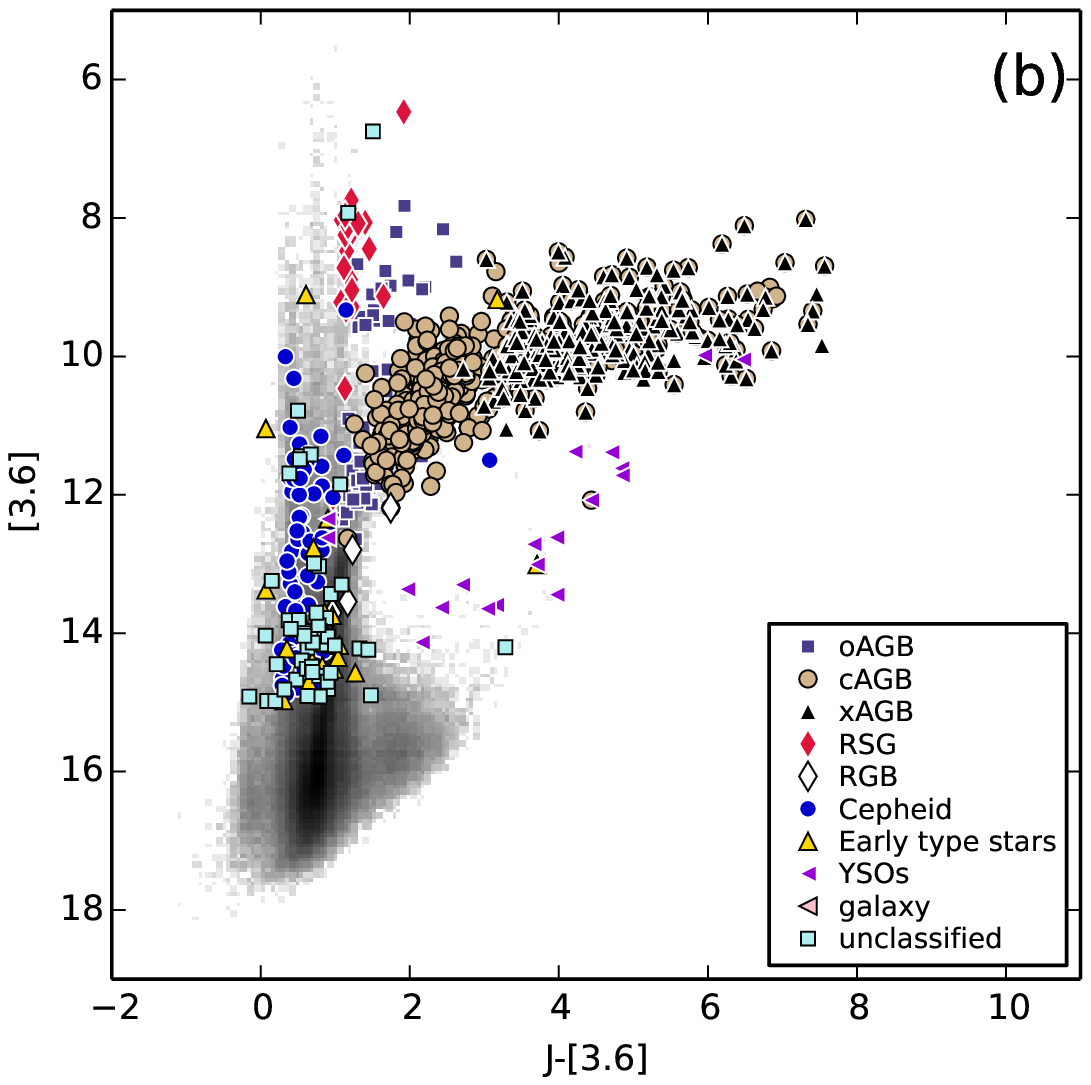} \\

                \includegraphics[width=.5\textwidth]{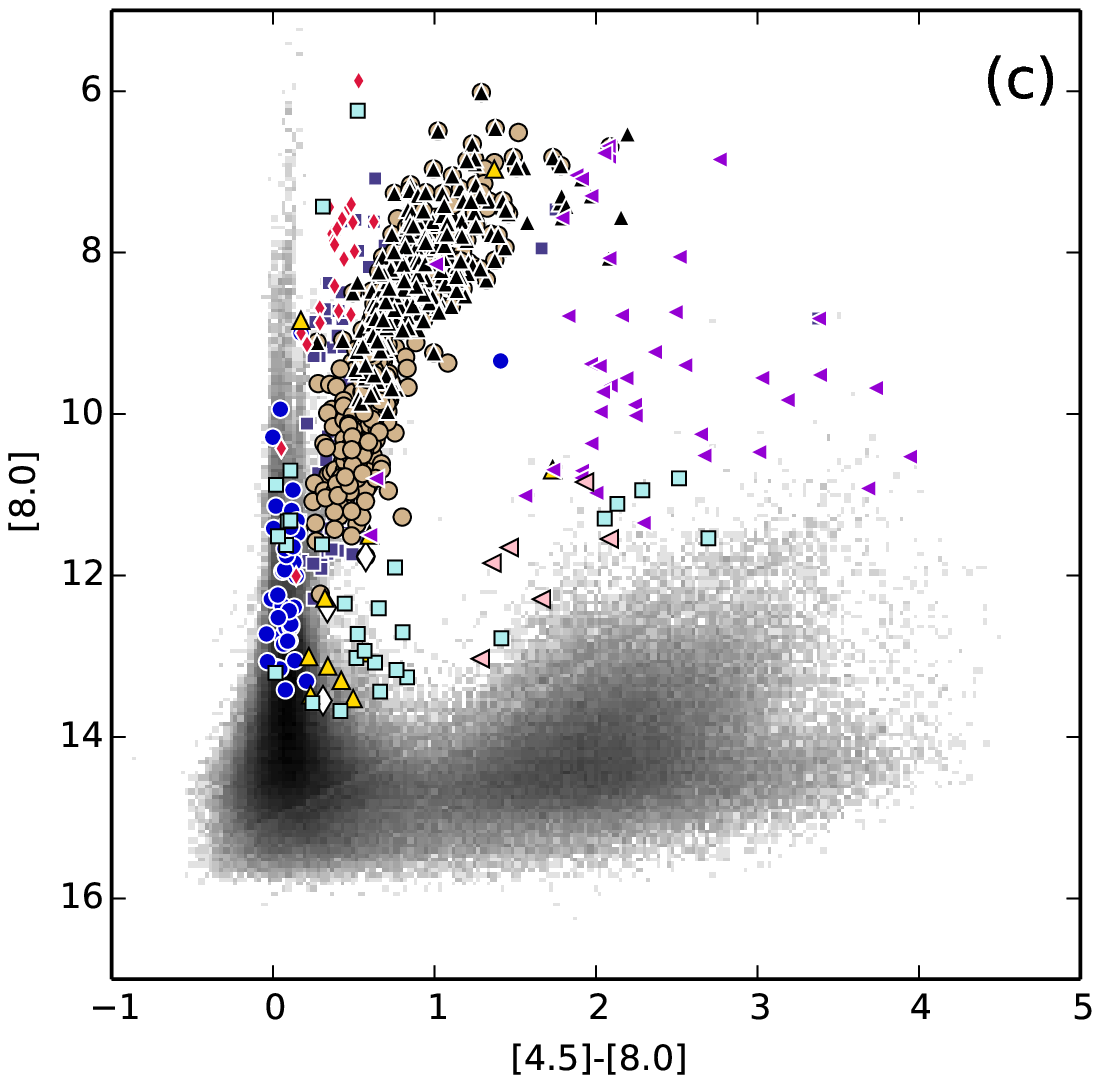}

                \includegraphics[width=.5\textwidth]{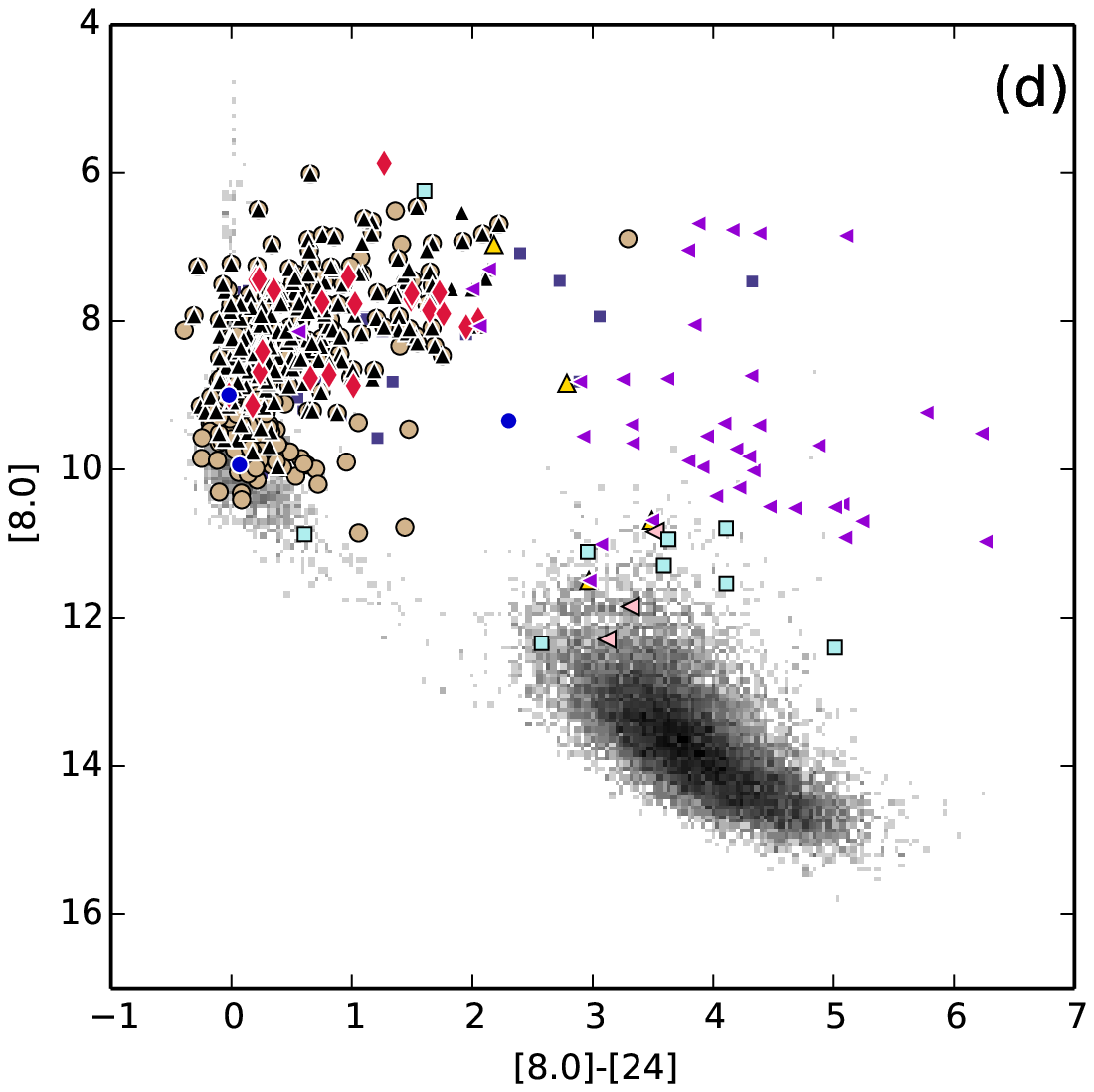}

  \end{tabular}
        \caption{The $K_s$ vs. $J-K_s$  (a), [3.6] vs. $J$-[3.6] (b), [8.0] vs. [4.5]-[8.0] (c), and [8.0] vs. [8.0]-[24] (d) CMDs showing the distribution of the variable star candidates with respect  to the entire population of sources from the SAGE-SMC IRAC Epoch 1 catalog (shown in grey as a Hess diagram).  CMDs in the top and bottom panel are examples of CMDs used to classify evolved stars (e.g., \citealp{boyer11}) and YSOs (e.g., \citealp{sewilo13}; \citealp{chen14}), respectively.  The symbols represent different classes of objects as indicated in the legends. Sources with multiple matches have all classifications plotted.}
        \label{f:fig6}
\end{figure}

In Figures~\ref{f:fig6}a-d, we have plotted our candidate infrared variables in a number of color-magnitude diagrams (CMDs), which allow a better understanding of the nature of these objects and can be used to search for anomalies (sources with an assigned class that show  unique spectral properties for sources of that type). In all plots, the entire SAGE-SMC Epoch 1 catalog is shown in greyscale as a source density plot (a Hess diagram).

Figure~\ref{f:fig6}a shows the near-IR CMD used for source selection by \citet{boyer11}. Indeed, in this diagram the evolved variable stars of different classes appear well separated, as a consequence of our source classification process heavily relying on the \citet{boyer11} catalog for evolved stars. In particular, the diagram demonstrates the neat color separation of extreme AGB stars from ``regular'' carbon stars (again a consequence of using the \citealt{boyer11} catalog for source classification): note, however, how many of these sources are still classified as C-stars by \citet{soszynski11}. Another prominent feature in this diagram is the vertical sequence of Cepheid stars (blue circles), which is well isolated from the redder evolved stars, and from the much bluer early type stars. The YSO sequence is also well separated and shows how sources with larger infrared excesses are also brighter in the $K_s$ band. Of particular interest among the outliers in this diagram is a very red  ($J - K_s \simeq 1.37$) source (SSTISAGE1C J005107.19-734133.3) classified as a Type II Cepheid in the OGLE-III database. This source is discussed in detail in Section~\ref{ss:redcepheid}. Figure \ref{f:fig6}a also shows a variable source (SSTISAGE1C J010921.65-712435.0) along the YSO sequence that is matched to both the YSO and C-rich AGB catalogs. The OGLE-III light curve for this variable is consistent with being a Mira, which would support its classification as a carbon star, despite its infrared colors being more consistent with the source being a YSO. The anomalous colors of this source could be explained by a photometric blend of two different sources, or by the presence of a circumstellar disk around a pulsating AGB star, which would mimic the spectral energy distribution of a YSO. Further analysis is required to confirm the nature of this source.

A similar separation between  different classes of SAGE-SMC infrared variables is shown in panels b and c of Figure \ref{f:fig6}. The low and intermediate mass evolved star sequence is dominated by (in order of increasing brightness and redder colors) RGBs, O-rich AGBs, C-rich AGBs, and extreme AGBs (most of them classified as having a C-rich chemistry). The RSGs have infrared colors similar to O-rich AGBs (due to the similar dust composition), but generally higher luminosity. Note an isolated group of O-rich AGBs with $[3.6] \la 10$, having a comparable brightness to RSGs, but moderately redder colors (by up to $J - [3.6] \simeq 1$~mag, indicating higher mass loss rate): these are likely candidate intermediate mass AGB stars undergoing HBB. In these diagrams, the YSOs are better isolated from the main sequence and the lower giant branch, but the brightest members of this group (likely high mass YSOs) still have infrared colors and magnitudes analogous to the reddest extreme AGB stars.

Figure~\ref{f:fig6}d is the CMD with the smallest number of sources due to the lower sensitivity of the MIPS 24~\micron{} band. However, it provides a better separation of the YSOs from all other classes, including background galaxies (the latter being concentrated in the large clump with $[8.0] \ga 12$ and $[8.0]-[24] \ga 2.5$). Of all the infrared variable active galaxies identified in our search, two are detected at 24~\micron{} and are located at the upper end of this clump.

We further investigate the infrared properties of candidate variables identified in our study by analyzing in detail their complete SEDs. This is discussed in the following sections.

\subsection{Spectral Energy Distribution}
\label{ss:sed}

\begin{figure}[h!]
  \centering
     \includegraphics{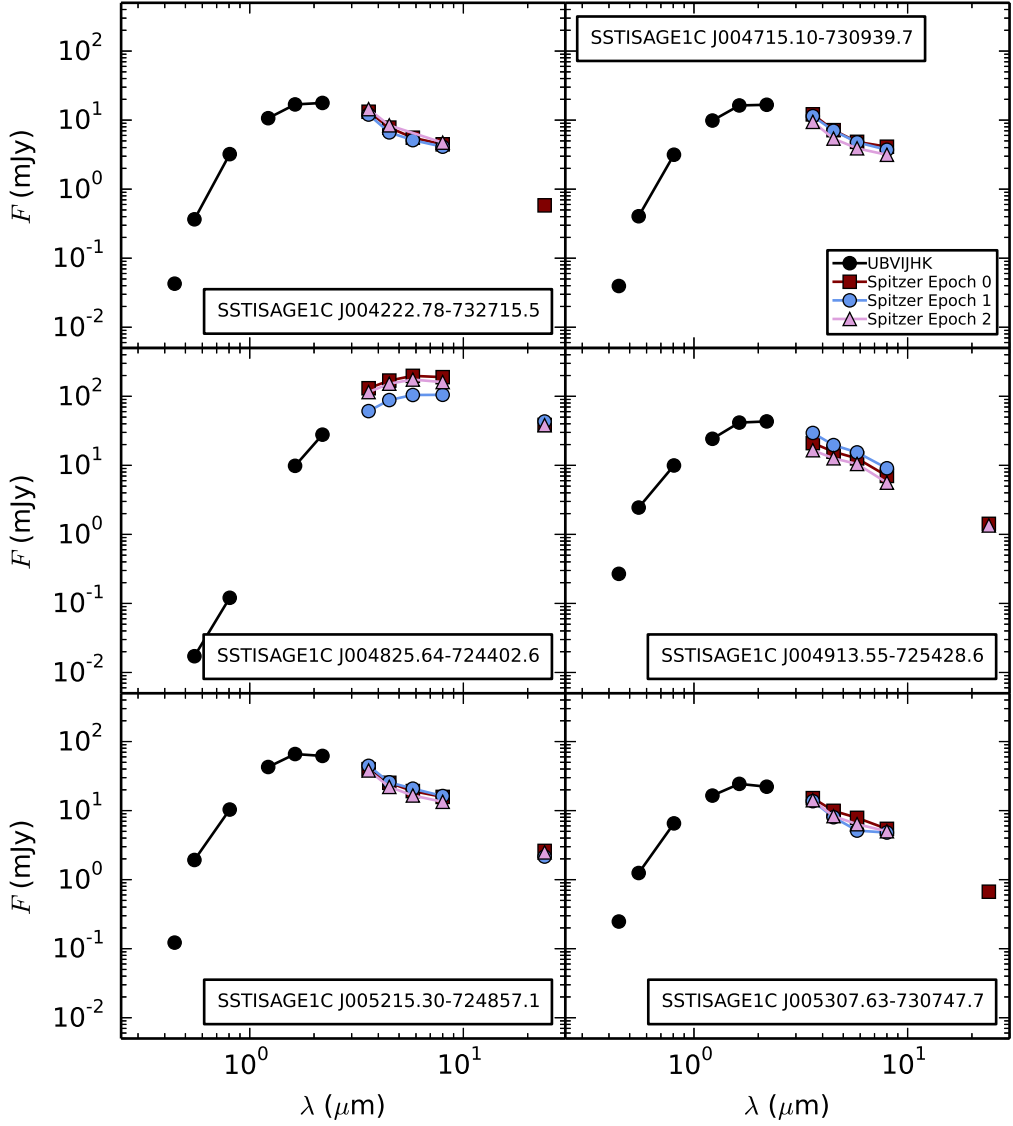}
  \caption{SEDs showing a representative sample of variable sources classified as C-rich AGB stars. The SAGE-SMC IRAC and MIPS 24 $\mu$m Epoch 0, Epoch 1, and Epoch 2 fluxes are plotted as red squares, blue circles, and pink triangles, respectively. The ancillary data from MCPS, OGLE-III, and 2MASS/6X2MASS surveys (see Section 2.1) are shown as black circles. The data obtained with the same instrument are connected by solid lines. }
\label{f:fig7}
\end{figure}

\begin{figure}[h!]
  \centering
      \includegraphics{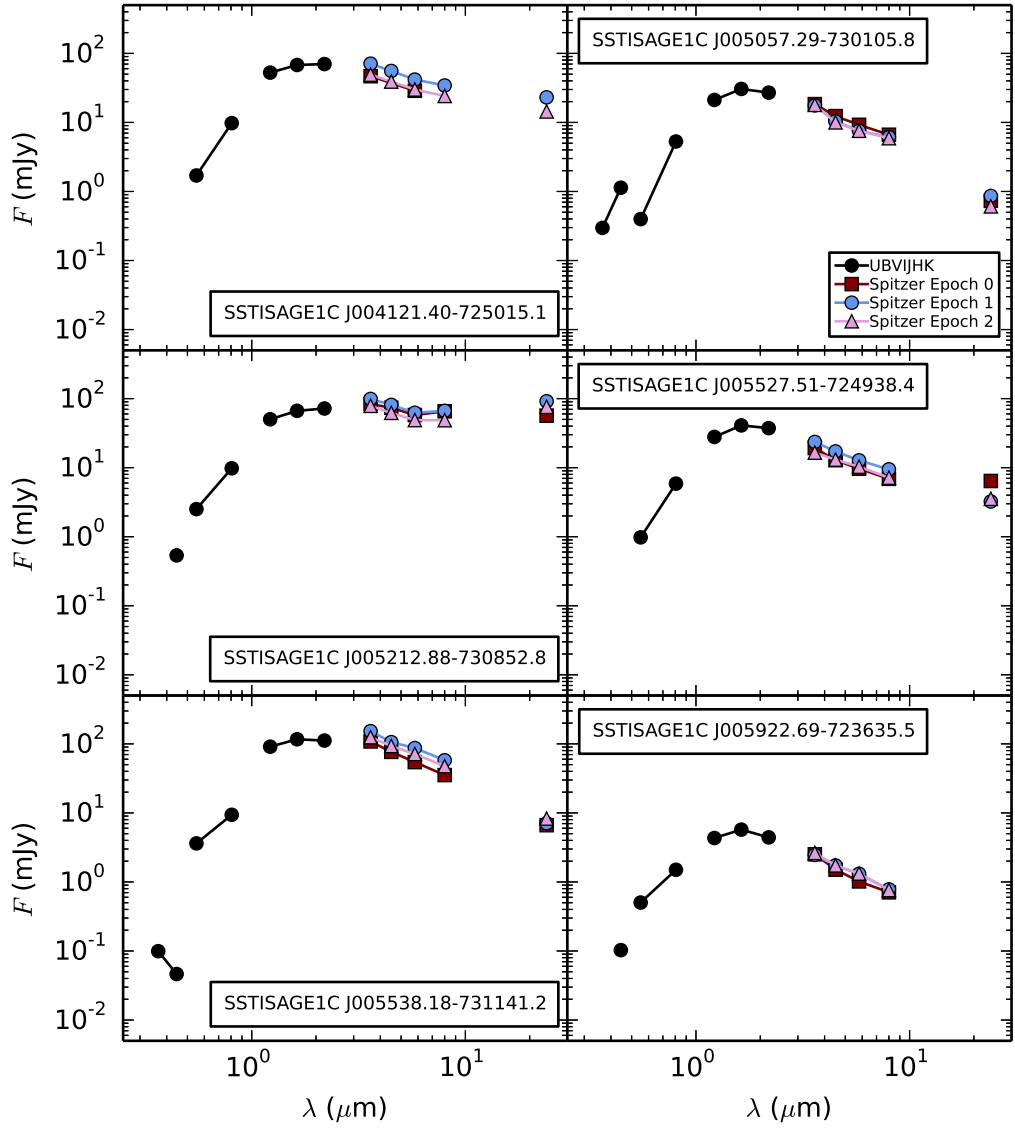}
  \caption{SEDs showing a representative sample of variable sources classified as O-rich AGB stars. The symbols and lines are the same as in Figure \ref{f:fig7}.}
\label{f:fig8}
\end{figure}

\begin{figure}[h!]
  \centering
      \includegraphics{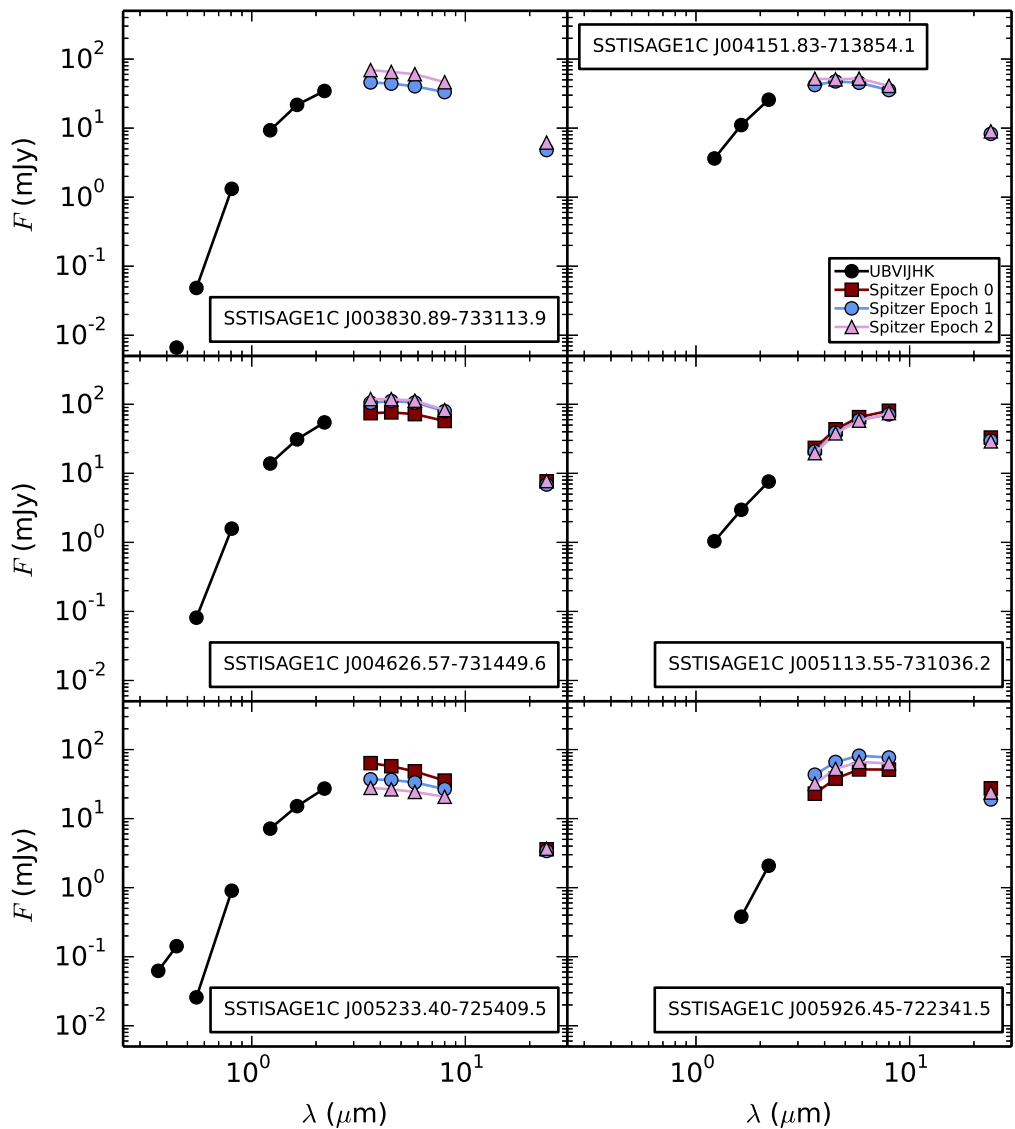}
  \caption{SEDs showing a representative sample of variable sources classified as extreme AGB stars. Most of these sources have a C-rich chemistry. The symbols and lines are the same as in Figure \ref{f:fig7}.}
\label{f:fig9}
\end{figure}

\begin{figure}[h!]
  \centering
     \includegraphics{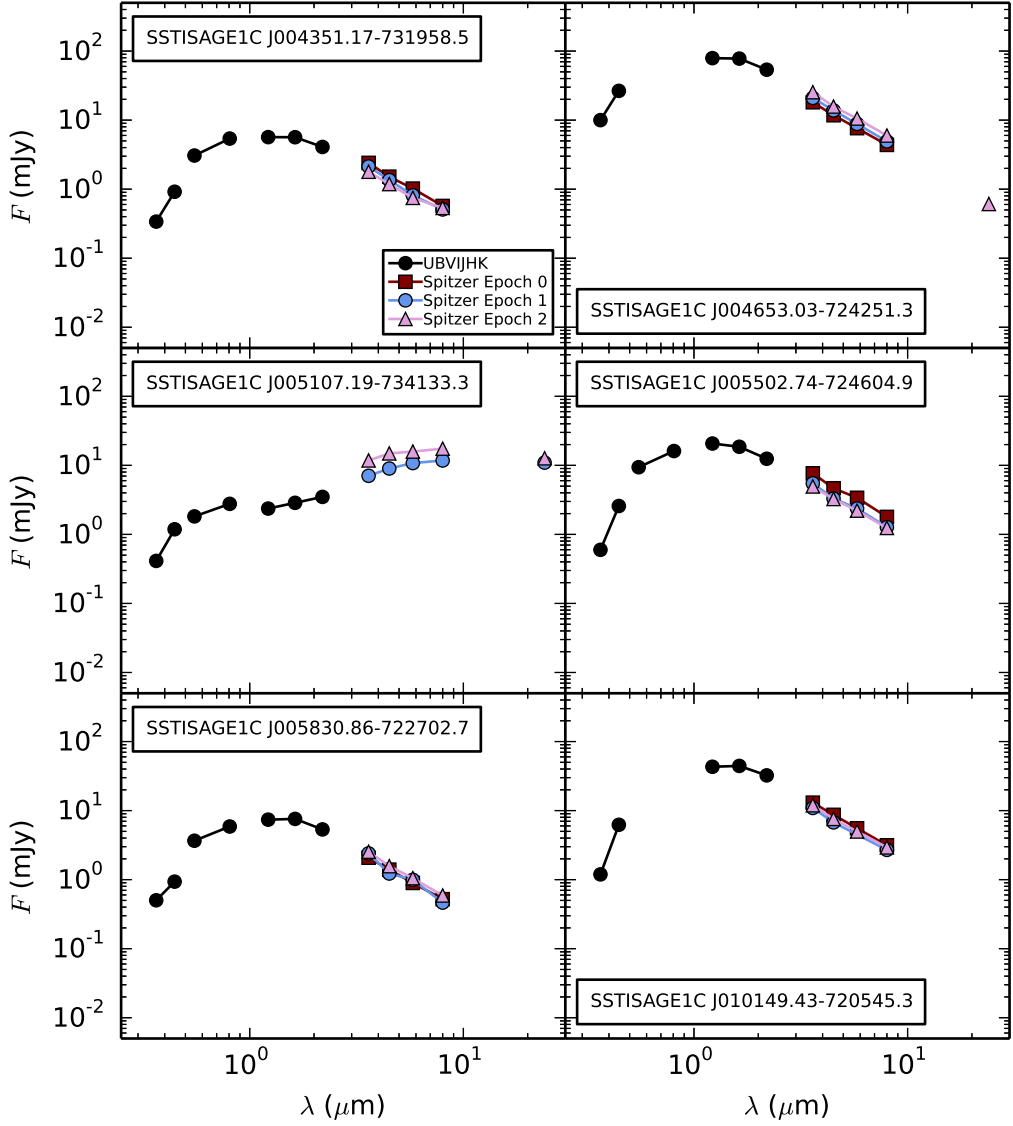}
  \caption{SEDs showing a representative sample of sources classified as Cepheid variable stars. Note that the very red source SSTISAGE1C J005107.19-734133.3 is not a Cepheid, but rather a post-AGB star with RV Tau variability (see Section~\ref{ss:redcepheid}). The symbols and lines are the same as in Figure~\ref{f:fig7}.}
\label{f:fig10}
\end{figure}

\begin{figure}[h!]
  \centering
      \includegraphics{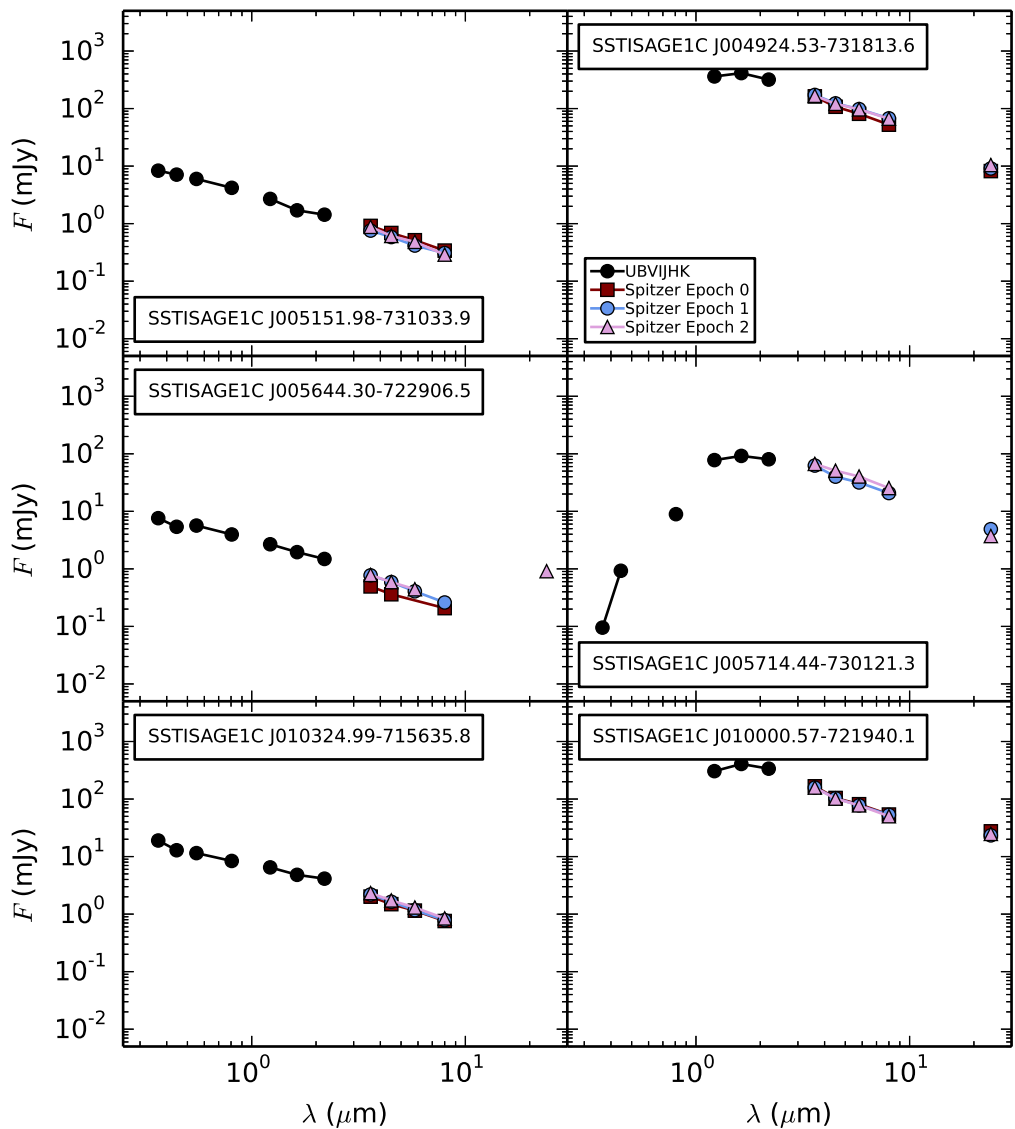}
  \caption{SEDs showing a representative sample of variable sources classified as massive stars.  SEDs of early type stars and RSGs are shown in the left and right panel, respectively. The symbols and lines are the same as in Figure~\ref{f:fig7}.}
\label{f:fig11}
\end{figure}

\begin{figure}[h!]
  \centering
      \includegraphics{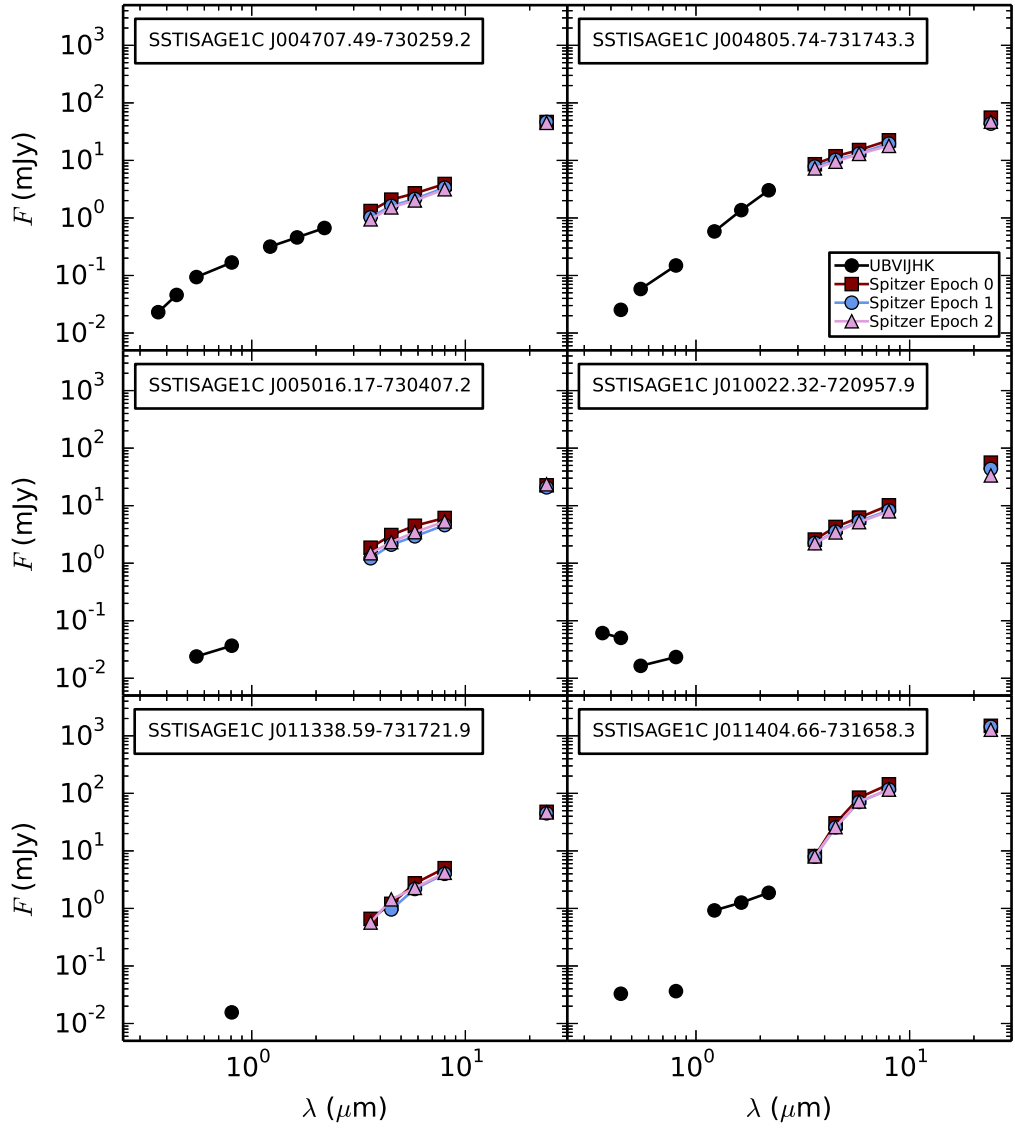}
  \caption{SEDs showing a representative sample of variable sources classified as YSOs. The symbols and lines are the same as in Figure \ref{f:fig7}.}
\label{f:fig12}
\end{figure}

\begin{figure}[h!]
  \centering
      \includegraphics{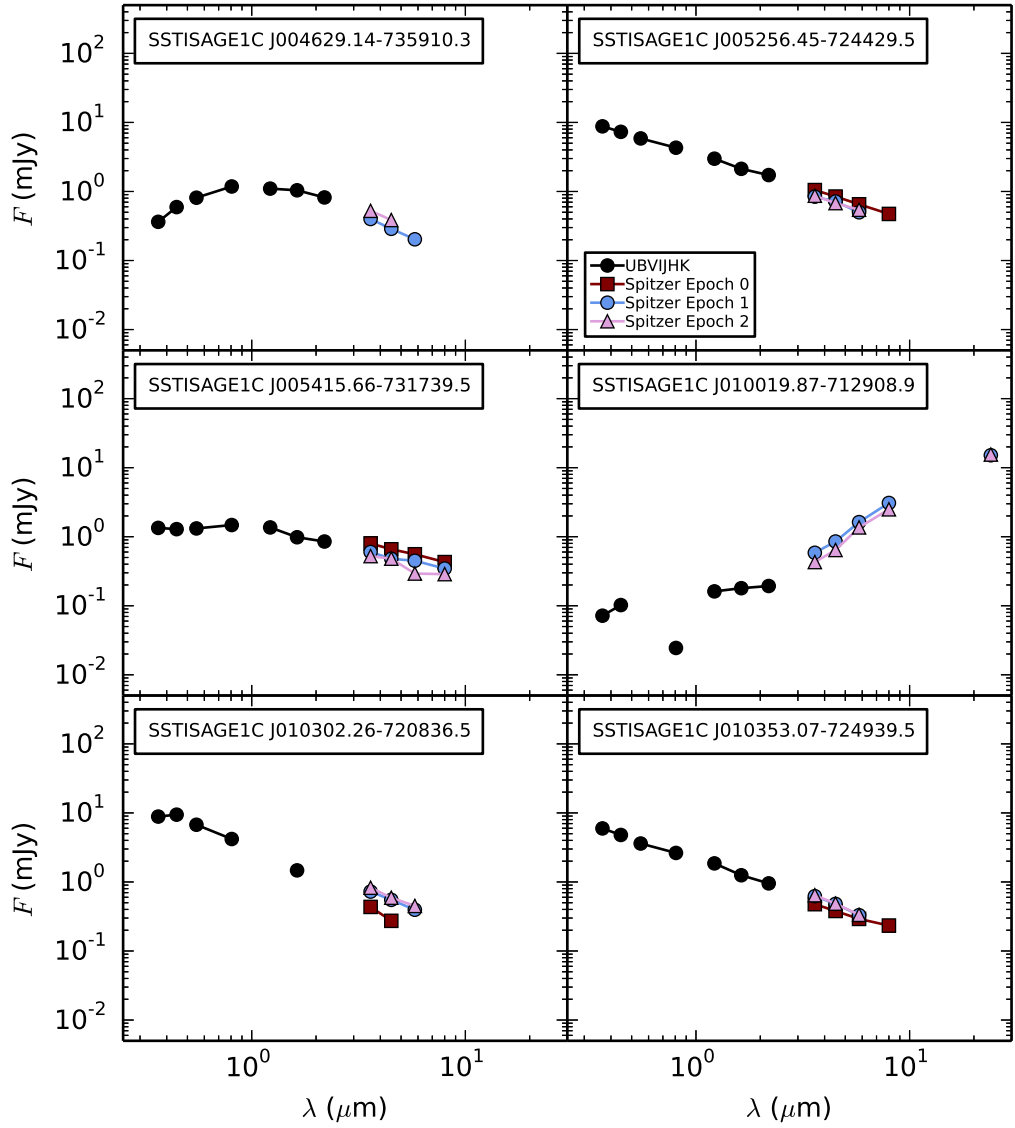}
  \caption{SEDs showing a representative sample of unclassified variable sources. The symbols and lines are the same as in Figure \ref{f:fig7}.}
\label{f:fig13}
\end{figure}

Figures~\ref{f:fig7} to \ref{f:fig13} show SEDs of a representative sample of variable sources of different types.
The {\it UB} (MCPS), {\it VI} (OGLE-III if available, MCPS otherwise), and {\it JHK$_{s}$} (2MASS/6X2MASS) data are plotted as black circles.  All three-epoch IRAC and MIPS 24 $\mu$m photometric data are plotted and color-coded as indicated in the plots. The data points from individual surveys (representing separate epochs of observations) are connected by solid lines.  The IRAC and MIPS 24 $\mu$m data points are not connected since the IRAC and MIPS data were not taken at the same time within each epoch. 

Figures~\ref{f:fig7}, \ref{f:fig8}, and \ref{f:fig9} show typical SEDs of evolved stars. The SEDs of O-rich and C-rich stars are very similar, both in the shape and in the amount of  flux variations observed in the IRAC and MIPS 24~\micron{} bands. The O-rich AGBs tend to have brighter 24~\micron{} fluxes, possibly due to the presence of the broad 18~\micron{} silicate band. The SEDs of the extreme AGB stars tend to peak at longer wavelengths.  Extreme AGBs are very bright at 24~\micron{} and some of them show very large variability in all IRAC and MIPS  bands.

Cepheid stars present typical SEDs of stellar photospheres with no evidence for an infrared excess (Figure~\ref{f:fig10}).  Among the massive stars in our sample, early type stars similarly show the characteristic SEDs of stellar photospheres, while RSGs have SEDs similar to the ones of O-rich AGBs, with generally small brightness changes in the IRAC and MIPS bands (see Figure~\ref{f:fig11}).  The SEDs of YSOs (Figure~\ref{f:fig12}) are the reddest in our sample, steeply increasing from optical to thermal infrared wavelengths.

Figure~\ref{f:fig13} shows SEDs of a representative sample of our unclassified sources. Most of the unclassified sources appear to be near or on the main sequence in Figure \ref{f:fig6}, panels a and b. However, Figure \ref{f:fig6}c shows that some of these sources appear to be slightly redder ($\sim$0.5 mag in $[4.5]-[8.0]$) than main sequence stars. It is possible that a fraction of the unclassified sources are older YSOs approaching the T Tauri phase, without protostellar envelopes and a vestigial disk. Some unclassified sources, however, have SEDs very similar to those of very red YSOs, and indeed are found in the same CMD space where variable sources with matches in the YSO catalogs are located.

\subsection{SSTISAGE1C J005212.88-730852.8: a Candidate SMC OH/IR Star}
\label{ss:ohirstar}

SSTISAGE1C J005212.88-730852.8 has one of the highest variability indices at 24~\micron{} in our sample: $V_{24} = -46.25$ between Epochs 0 and 1. The negative sign indicates that the source has been declining in brightness between November 2004 (S$^3$MC) and September 2007 (SAGE-SMC Epoch 1). The variability index in the 24~\micron{}  band between Epochs 0 and 2 (June 2008) is less negative ($V_{24} = -30.38$), suggesting that the source has been brightening in the 9 months between the two SAGE-SMC MIPS 24~\micron{} epochs. This is confirmed by the variability index we measure in all IRAC bands, which is positive between Epochs 1 and 2, indicating that the source has continued brightening in the three months between June and September 2008 (when the IRAC SAGE-SMC data were collected).

SSTISAGE1C J005212.88-730852.8 is classified by both \citet{boyer11} and \citet{soszynski11} as an O-rich AGB star. The OGLE-III catalog, in particular, classifies the source as a Mira with a long secondary period of 1,453 days, favoring its classification as an AGB star rather than a RSG (the latter being typically a Semiregular variable). The O-rich chemistry is confirmed by \citet{vanloon08}, who note that in the past the star has been confused with a carbon star. The optical spectral classification by \citet{blanco80} as an M6 giant, however, is in agreement with the \emph{Spitzer} and ISO spectra showing a silicate feature in emission \citep{cioni03}. The source is peculiar also because it is the only stellar object in our catalog that has a match in the \citet{wong12} catalog of radio sources: the star is detected in the 6~cm band, which is known for the presence of OH masers (the source is not detected in the 3~cm band, as expected if the emission is not radio continuum). These characteristics suggest that SSTISAGE1C J005212.88-730852.8 may be an OH/IR star: if confirmed, this could be the first OH/IR star in the SMC (none are currently known, according to \citealt{vanloon12}).

The source is very bright ($[3.6] = 8.63$ mag and $[24] = 4.73$ mag), suggesting that it may be one of the intermediate mass O-rich AGBs undergoing HBB. It is also very red ($[8.0] - [24] = 2.73$ mag), indicative of a large amount of cold circumstellar dust. In fact, the SED of this source (see Figure~\ref{f:fig8}), is quite similar to the flat, red spectra of extreme AGBs. The spectral characteristics suggest that SSTISAGE1C J005212.88-730852.8 may be one of the O-rich AGB stars with highest mass loss rate in the SMC, comparable to the one typically observed in the C-rich extreme AGB stars. Finding sources of this kind is important to explain the overall dust budget in the Magellanic Clouds. As mentioned in Section~\ref{ss:evolvedstars}, the large dust producers in both the LMC and SMC tend to be extreme carbon stars \citep{boyer12, riebel12}, making it difficult to explain the average composition of the Magellanic Clouds' ISM (which has similar fraction of O-rich and C-rich dust). Stars like SSTISAGE1C J005212.88-730852.8 may help to reduce the observed deficit in silicate dust production.

\subsection{SSTISAGE1C J005107.19-734133.3: an RV Tau Variable}
\label{ss:redcepheid}

SSTISAGE1C J005107.19-734133.3 is a very red star classified as an extreme AGB by \citet{boyer11} and as a Type II Cepheid (with a period of 39.52~days) in the OGLE-III database \citep{soszynski10b}. As noted in Section \ref{ss:cepheids}, the star is much redder than all other Cepheids, and its SED is dominated by a strong infrared excess indicative of copious amounts of circumstellar dust. The absolute variability index of this source at 24~\micron{} between Epochs 1 and 2 is larger than 5, the highest for all sources in the OGLE-III Cepheid class, and larger than 7 in all IRAC bands.

SSTISAGE1C J005107.19-734133.3 is in fact among the post-AGB stars spectrally classified by \citet{kamath14} in the SMC. In their analysis, \citet{kamath14} confirms that the source shows an RV Tau type variability with alternating shallow and deep minima, and an infrared excess consistent with the presence of a circumbinary disk, as common in post-AGB stars. As noted by \citet{soszynski10b}, RV Tau variables are matched with Type II Cepheids in the OGLE-III automatic classification light curve scheme. Visual inspection of the optical spectra of this source suggests $s$-process enrichment, evidence that the star has recently gone through the third dredge-up, and thus may have only recently left the AGB phase.

\section{Conclusions}
\label{s:conclusions}

We have identified and characterized a population of candidate infrared variables in the SMC using the data from the  \emph{Spitzer} SAGE-SMC and S$^{3}$MC surveys.  We found 814 sources showing variability in two or more adjacent mid-infrared bands. The variable source population is dominated by evolved stars ($\sim$75\%), specifically C-rich AGBs ($\sim$60\%). Around 10\% of our variable sources are O-rich AGB stars, with one variable star candidate thought to be a rare OH/IR source. We find that $\sim$3\% and $<$1\% of our variable sources are RSG and RGB stars, respectively.  Our list of candidate infrared variables include a small, but significant population of Cepheid variables ($\sim$9\%), as well as early type variable stars ($\sim$3\%), and variable YSOs  ($\sim$5\%). A little over 1\% of our sources are background galaxies that vary in the infrared. Around 7\% of sources in our catalog are currently unclassified, but may be further examined through future studies.

\acknowledgments {E. Polsdofer is thankful to the Association of Universities for Research in Astronomy for their initial support and funding of her summer internship program.
 
E. Polsdofer and M. Meixner are grateful for support from NASA NAG5-12595.

This work is based on observations made with the {\it Spitzer} Space Telescope, which is operated by the Jet Propulsion Laboratory, California Institute of Technology under a contract with NASA.

This publication makes use of data products from the Two Micron All Sky Survey, which is a joint project of the University of Massachusetts and the Infrared Processing and Analysis Center/California Institute of Technology, funded by the National Aeronautics and Space Administration and the National Science Foundation.}


\clearpage
\clearpage

\begin{deluxetable}{clllc}
\tablecaption{SAGE-SMC and S$^3$MC Observing Dates and Survey Areas}
\tabletypesize{\small}
\tablewidth{0pt}
\tablecolumns{5}
\tablehead{ 
\colhead{Epoch} & \colhead{Survey name} & \colhead{IRAC observation date} & \colhead{MIPS observation date}  & \colhead{Sky area (deg$^2$)}}
\startdata
0 		& S$^3$MC 		 & 2005 May 7-9 		& 2004 November 6-8 			& $\sim$3.7 \\
1 		& SAGE-SMC 			& 2008 June 12-19 		& 2007 September 17-25 		& $\sim$30 \\
2 		& SAGE-SMC 		& 2008 September 15-23 		& 2008 June 25-28 			& $\sim$30
\enddata
\label{t:tab1}
\end{deluxetable}

\clearpage

\begin{deluxetable}{lrrrcc}
\tablecaption{Variability Criteria and Source Counts}
\tabletypesize{\small}
\tablewidth{0pt}
\tablecolumns{6}
\tablehead{ 
\colhead{Criteria} & \colhead{3.6 $\mu$m} & \colhead{4.5 $\mu$m} & \colhead{5.8 $\mu$m} & \colhead{8.0 $\mu$m} & \colhead{24 $\mu$m}}
\startdata
\cutinhead{\bf{Epochs 0 and 1}}
\# of sources with valid fluxes & 169424	& 120190 & 27052 & 10172 & 1425 \\
\# with $|$v band $| >$ 3 & 1939 & 1708 & 570 & 409 & 189 \\
\% with $|$v band $| >$ 3 & 1.14 & 1.42 & 2.11 & 4.03 & 13.26 \\
\# of sources with valid fluxes &  &  &  &  &  \\
in neighboring bands & 115308$\rm{^{a}}$ & 116101$\rm{^{b}}$ & 23885$\rm{^{c}}$ & 11294$\rm{^{d}}$ & 1083$\rm{^{e}}$ \\
\# meet variability criterion & 194 & 407 & 466 & 617 & 623 \\
\% meet variability criterion & 0.17 & 0.35 & 1.95 & 5.46 & 57.53 \\

\cutinhead{\bf{Epochs 0 and 2}}
\# of sources with valid fluxes & 175714 & 123027 & 27066 & 12129 & 1551 \\
\# with $|$v band $| >$ 3 & 1963 & 1742 & 571 & 421 & 189 \\
\% with $|$v band $| >$ 3 & 1.12 & 1.42 & 2.11 & 3.47 & 12.19 \\
\# of sources with valid fluxes &  &  &  &  &  \\
in neighboring bands & 112696$\rm{^{a}}$ & 113568$\rm{^{b}}$ & 23917$\rm{^{c}}$ & 9552$\rm{^{d}}$ & 976$\rm{^{e}}$ \\
\# meet variability criterion & 196 & 411 & 483 & 645 & 646 \\
\% meet variability criterion & 0.17 & 0.36 & 2.02 & 6.75 & 66.19 \\

\cutinhead{\bf{Epochs 1 and 2}}
\# of sources with valid fluxes & 812930 & 551722 & 87886 & 36444 & 6544 \\
\# with $|$v band $| >$ 3 & 3561 & 3252 & 582 & 435 & 277 \\
\% with $|$v band $| >$ 3 & 0.44 & 0.59 & 0.66 & 1.19 & 4.23 \\
\# of sources with valid fluxes &  &  &  &  &  \\
in neighboring bands & 496470$\rm{^{a}}$ & 500512$\rm{^{b}}$ & 81995$\rm{^{c}}$ & 32302$\rm{^{d}}$ & 4092$\rm{^{e}}$ \\
\# meet variability criterion & 446 & 446 & 390 & 374 & 227 \\
\% meet variability criterion & 0.09 & 0.09 & 0.48 & 1.16 & 5.55 \\
\enddata
\tablenotetext{a}{Valid fluxes at 3.6 and 4.5 $\mu$m.}
\tablenotetext{b}{Valid fluxes at 3.6 and 4.5 $\mu$m or at 4.5 and 5.8 $\mu$m.}
\tablenotetext{c}{Valid fluxes at 4.5 and 5.8 $\mu$m or at 5.8 and 8.0 $\mu$m.}
\tablenotetext{d}{Valid fluxes at 5.8 and 8.0 $\mu$m.}
\tablenotetext{e}{Valid fluxes at 8.0 and 24 $\mu$m.}
\label{t:tab2}
\end{deluxetable}

\clearpage

\begin{deluxetable}{lll}
\tabletypesize{\small}
\centering 
\tablecaption{Properties of the SAGE-SMC Infrared Variables}
\tablehead{ 
\colhead{Column} & \colhead{Name} & \colhead{Description}
}
\startdata 
1 & varid 		& The identification name of the variable star candidates \\
2 & R.A. (J2000) & 		Right Ascension, J2000 (deg) \\
3 & Decl. (J2000) & 		Declination, J2000 (deg) \\
4-6 & IRACdesignations\_e\textit{i} 		& SAGE-SMC IRAC designations for Epoch 0 (\textit{i}=0), \\
 & 		& Epoch 1 (\textit{i}=1), and Epoch 2 (\textit{i}=2) \\
7-9 & MIPSdesignations\_e\textit{i} 		& SAGE-SMC MIPS designations for Epoch 0 (\textit{i}=0), \\
 & 		& Epoch 1 (\textit{i}=1), and Epoch 2 (\textit{i}=2) \\
10-12 & IRACdistance\_\textit{i} 		& Distance between SAGE-SMC IRAC Epochs 0 and 1 (\textit{i}=01), \\
 & 		& Epochs 0 and 2 (\textit{i}=02), Epochs 1 and 2 (\textit{i}=12) (arcsec) \\
13-15 & MIPSdistance\_e\textit{i}	 	& Distance between SAGE-SMC IRAC and MIPS for Epoch 0 (\textit{i}=0), \\
 & 		& Epoch 1 (\textit{i}=1), and Epoch 2 (\textit{i}=2) \\
16 & E01\tablenotemark{a}		& Variability flag between Epochs 0 and 1 \\
17 & E02\tablenotemark{a} 		& Variability flag between Epochs 0 and 2 \\
18 & E12\tablenotemark{a} 		& Variability flag between Epochs 1 and 2 \\
19-21 & classification\textit{j} 			& Classification of sources (\textit{j}=1-3) \\
22-26 & V\textit{k}\_01		& Variability indices between Epochs 0 and 1 for the \\
 & 		& IRAC 3.6, 4.5, 5.8,  8.0 $\mu$m (\textit{k}=1-4) and MIPS 24 $\mu$m (\textit{k}=5) bands. \\
27-31 & V\textit{k}\_02 		& Variability indices between Epochs 0 and 2 for the \\
 & 		& IRAC 3.6, 4.5, 5.8, 8.0 $\mu$m (\textit{k}=1-4) and MIPS 24 $\mu$m (\textit{k}=5) bands. \\
32-36 & V\textit{k}\_12 		& Variability indices between Epochs 1 and 2 for the \\
 & 		& IRAC 3.6, 4.5, 5.8,  8.0 $\mu$m (\textit{k}=1-4) and MIPS 24 $\mu$m (\textit{k}=5) bands. \\
37-46 & flux\textit{k}, dflux\textit{k} 		& Fluxes and flux uncertainties (mJy) for MCPS \textit{UBVI}\tablenotemark{b} (\textit{k}=1-8), \\
 & 		& OGLE-III \textit{VI}\tablenotemark{c,d}~  (\textit{k}=9-10),  2MASS \textit{JHK$_s$} (\textit{k}=11-16) \\
47-62 & e0\_flux\textit{k}, e0\_dflux\textit{k} 		& Fluxes and flux uncertainties (mJy) for Epoch 0:\\
 & 		& \textit{Spitzer} IRAC 3.6, 4.5, 5.8, 8.0 $\mu$m (\textit{k}=1-4) and MIPS 24 $\mu$m (\textit{k}=5) bands. \\
63-72 & e1\_flux\textit{k}, e1\_dflux\textit{k} 		& Fluxes and flux uncertainties (mJy) for Epoch 1:\\
 & 		& \textit{Spitzer} IRAC 3.6, 4.5, 5.8, 8.0 $\mu$m (\textit{k}=1-4) and MIPS 24 $\mu$m (\textit{k}=5) bands. \\
73-82 & e2\_flux\textit{k}, e2\_dflux\textit{k} 		& Fluxes and flux uncertainties (mJy) for Epoch 2:\\
& 		& \textit{Spitzer} IRAC 3.6, 4.5, 5.8, 8.0 $\mu$m (\textit{k}=1-4) and MIPS 24 $\mu$m (\textit{k}=5) bands. \\
83-96 & mag\textit{k}, dmag\textit{k} 		& Magnitudes and magnitude uncertainties (mag) \\
 & 		& for MCPS \textit{UBVI} (\textit{k}=1-8) and 2MASS \textit{JHK$_s$} (\textit{k}=9-14) \\
97-106 & e0\_mag\textit{k}, e0\_dmag\textit{k} 		& Magnitudes and magnitude uncertainties (mag) for Epoch 0:\\
 & 		& \textit{Spitzer} IRAC 3.6, 4.5, 5.8, 8.0 $\mu$m (\textit{k}=1-4) and MIPS 24 $\mu$m (\textit{k}=5) bands.\\
107-116 & e1\_mag\textit{k}, e1\_dmag\textit{k} 		& Magnitudes and magnitude uncertainties (mag) for Epoch 1:\\
 & 		& \textit{Spitzer} IRAC 3.6, 4.5, 5.8, 8.0 $\mu$m (\textit{k}=1-4) and MIPS 24 $\mu$m (\textit{k}=5) bands. \\
117-126 & e2\_mag\textit{k}, e2\_dmag\textit{k} 		& Magnitudes and magnitude uncertainties (mag) for Epoch 2:\\
 & 		& \textit{Spitzer} IRAC 3.6, 4.5, 5.8, 8.0 $\mu$m (\textit{k}=1-4) and MIPS 24 $\mu$m (\textit{k}=5) bands. \\
\enddata 
\tablenotetext{a}{Flags indicating if a given source has been identified as a candidate infrared variable based on the data from two epochs
(E01 - Epochs 0 and 1; E02 - Epochs 0 and 2; E12 - Epochs 1 and 2):  1 -- a source is a variable, i.e., its absolute variability index is larger than 3 in at least 2 consecutive bands; 0 -- a source does not vary or there is not enough photometric data available. }
\tablenotetext{b}{The MCPS catalog provides magnitudes only. The \textit{U}-, \textit{B}-, \textit{V}-, and \textit{I}-band fluxes and flux uncertainties
were derived using the zero magnitude flux of 1790, 4063, 3636, and 2416 Jy, respectively.}
\tablenotetext{c}{No photometric uncertainties are available for the OGLE-III data.}
\tablenotetext{d}{The OGLE-III catalog provides magnitudes only. The \textit{V}- and \textit{I}-band fluxes were derived using the zero magnitude flux of 3636 Jy and 2416 Jy, respectively.}
\label{t:tab3}
\end{deluxetable}


\clearpage

\begin{deluxetable}{lcc}
\tabletypesize{\small}
\centering
\tablecaption{Catalog Matching Results}
\tablehead{ 
\colhead{Literature} & \colhead{Matched contents} & \colhead{\# of sources}
}
\startdata
\cite{bolatto07} 		& YSOs 		& 13 \\
\cite{bonanos10} 		& Massive stars$\rm{^{a}}$ 		& 43 \\
\cite{boyer11} 		& 	Evolved stars$\rm{^{b,f}}$ 		& 600 \\
\cite{carlson11} 		& YSOs 		& 3 \\
G. Jacoby (private communication) 		& PNe 		& 1 \\
\cite{kozlowski11} 		& AGN$\rm{^{c}}$ 		& 8 \\
\cite{kozlowski13} 		& QSOs$\rm{^{c}}$ & 1 \\
\cite{oliveira11} 		& YSOs 		& 1 \\
\cite{oliveira13}	    & YSOs 		& 4 \\
\cite{seale14} 		& YSOs 		& 10 \\
\cite{sewilo13} 		& YSOs 		& 44 \\
\cite{sheets13} 		& dustyOB$\rm{^{d}}$ 		& 1 \\
\cite{simon07} 		& YSOs 		& 3 \\
\cite{soszynski10a} 		& Classic Cepheids$\rm{^{e}}$ 		& 67 \\
\cite{soszynski10b} 		& Type II Cepheids$\rm{^{e}}$ 		& 3 \\
\cite{soszynski11} 		& Evolved stars$\rm{^{f}}$		& 539 \\
\cite{sturm13} 		& X-ray sources 		& 6 \\
\cite{vanloon08} 		& YSOs 		& 1\\
\cite{vanloon10} 		& YSOs 		& 1 \\
\cite{wong12} 		& Radio 		& 1 
\enddata
\tablenotetext{a}{Includes early type and RSG stars.}
\tablenotetext{b}{Includes cAGB, xAGB, oAGB, RGB and RSG stars.}\\
\tablenotetext{c}{Reported as a background galaxy.}
\tablenotetext{d}{Reported as an early type star.}
\tablenotetext{e}{Reported as a Cepheid.}
\tablenotetext{f}{Includes cAGB, xAGB, oAGB and RGB stars.}
\label{t:tab4}
\end{deluxetable}

\clearpage

\end{document}